\newif\ifjmr

\newif\ifreview

\newcommand{\papertitle}{Correlated Dynamics in Marketing Sensitivities}

\ifjmr
    \documentclass[12pt]{article}
    
\AtBeginEnvironment{appendices}{\crefalias{section}{appendix}\crefalias{subsection}{appendix}\crefalias{subsubsection}{appendix}}

\usepackage{natbib}
\usepackage[affil-it]{authblk}
\usepackage{amsmath, amssymb, mathtools, amsfonts, amsthm, verbatim, nicefrac, bm, bbm, fancyhdr, array, listings, xcolor, enumitem, thmtools, subcaption, pdflscape, pgffor, csquotes, listings, lstbayes, longtable, soul, booktabs, tablefootnote, makecell, multirow, graphicx, color, xcolor, pdflscape, adjustbox}

\definecolor{main}{rgb}{0.05,0.25,0.56}
\definecolor{dark-red}{rgb}{0.4,0.15,0.15}
\definecolor{dark-blue}{rgb}{0.05,0.25,0.56}
\definecolor{light-blue}{rgb}{.2,1,1}
\definecolor{medium-blue}{rgb}{0,0,0.5}
\definecolor{dark-green}{rgb}{0,.5,0}
\usepackage[colorlinks, pagebackref=false]{hyperref}
\usepackage{caption}
\usepackage[nameinlink]{cleveref}
\hypersetup{
    citecolor={dark-blue},
    bookmarksnumbered=true,
    urlcolor=dark-red,
    linkcolor=dark-blue
}
\creflabelformat{equation}{#2#1#3}
\usepackage{mathpazo}
\usepackage{helvet}
\usepackage{colortbl}

\newcolumntype{L}[1]{>{\raggedright\let\newline\\\arraybackslash\hspace{0pt}}p{#1}}
\newcolumntype{C}[1]{>{\centering\let\newline\\\arraybackslash\hspace{0pt}}m{#1}}
\newcolumntype{R}[1]{>{\raggedleft\let\newline\\\arraybackslash\hspace{0pt}}m{#1}}
\newcolumntype{H}{>{\setbox0=\hbox\bgroup}c<{\egroup}@{}}

\usepackage{titlesec}
\usepackage[margin=1in]{geometry}
\usepackage[nodisplayskipstretch]{setspace}
\usepackage{ragged2e}
\usepackage{etoolbox}

\usepackage{fancyhdr}
\titleformat{\section}{\centering \bf \uppercase}{}{0pt}{}
\titleformat{\subsection}{\bf \em}{}{0pt}{}
\titleformat{\subsubsection}[runin]{\em}{}{0pt}{}[.~~]
\titleformat{\paragraph}[runin]{\em}{}{0pt}{}[.~~]
\titlespacing{\section}{0pt}{24pt}{24pt}
\titlespacing{\subsection}{0pt}{12pt}{12pt}

\RaggedRight
\renewcommand{\sffamily}{}

\usepackage{mathpazo}

    \title{\papertitle}
\else
    \documentclass[11pt]{article}
\usepackage[pdftex,letterpaper,left=1in,top=1in,bottom=1in,right=1in]{geometry}
\usepackage{afterpage, setspace} 
\usepackage[titletoc,title]{appendix}
\usepackage[dotinlabels]{titletoc}
\usepackage{titlesec, marginnote, ftnxtra, todonotes, minitoc}
\usepackage[round,longnamesfirst]{natbib}
\setcitestyle{aysep={}} 
\usepackage[nottoc]{tocbibind} 
\usepackage{multibib}
\newcites{latex}{References (Appendix)} 
\AtBeginEnvironment{appendices}{\crefalias{section}{appendix}\crefalias{subsection}{appendix}\crefalias{subsubsection}{appendix}}

\usepackage{amsmath, amssymb, mathtools, amsfonts, amsthm, verbatim, nicefrac, bm, bbm, fancyhdr, array, listings, xcolor, enumitem, thmtools, subcaption, pdflscape, pgffor, listings, lstbayes, longtable, soul}
\usepackage{booktabs, tablefootnote, makecell, multirow}
\usepackage[referable]{threeparttablex}
\setlist{noitemsep,parsep=6pt,partopsep=0pt,topsep=0pt}
\usepackage{graphicx, color, xcolor}
\usepackage{pdflscape}
\usepackage{adjustbox}
\usepackage{ragged2e}

\definecolor{main}{rgb}{0.05,0.25,0.56}
\definecolor{dark-red}{rgb}{0.4,0.15,0.15}
\definecolor{dark-blue}{rgb}{0.05,0.25,0.56}
\definecolor{light-blue}{rgb}{.2,1,1}
\definecolor{medium-blue}{rgb}{0,0,0.5}
\definecolor{dark-green}{rgb}{0,.5,0}
\usepackage[colorlinks, pagebackref=false]{hyperref}
\usepackage{caption}
\usepackage[nameinlink]{cleveref}
\hypersetup{
    citecolor={dark-blue},
    bookmarksnumbered=true,
    urlcolor=dark-red,
    linkcolor=dark-blue
}

\creflabelformat{equation}{#2#1#3}
\usepackage{mathpazo}
\usepackage[margin=15pt,font=small,labelfont={bf,sf}]{caption}
\usepackage{helvet}
\usepackage{sectsty}
\allsectionsfont{\sffamily\color{dark-blue}}
\usepackage{colortbl}

\newcolumntype{L}[1]{>{\raggedright\let\newline\\\arraybackslash\hspace{0pt}}p{#1}}
\newcolumntype{C}[1]{>{\centering\let\newline\\\arraybackslash\hspace{0pt}}m{#1}}
\newcolumntype{R}[1]{>{\raggedleft\let\newline\\\arraybackslash\hspace{0pt}}m{#1}}
\newcolumntype{H}{>{\setbox0=\hbox\bgroup}c<{\egroup}@{}}

    \title{\textbf{\sffamily\color{dark-blue} \papertitle}}
\fi

\ifreview
    \author{Authors masked for review.}
\else
    \author{Ryan Dew, Yuhao Fan\footnote{Ryan Dew is the corresponding author (\texttt{ryandew@wharton.upenn.edu}). He is an Assistant Professor of Marketing at the Wharton School, University of Pennsylvania. Yuhao Fan is a Data Scientist at URBN. Special thanks to Kevin Liu, Sarah Ye, Zhangyi Fan, and Nikhil Kona for excellent research assistance. Thanks, too, to Zhenling Jiang and Christophe Van den Bulte for helpful comments.}}
\fi

\begin{document}

\maketitle

\begingroup
\justifying
\begin{abstract}
\noindent Understanding individual customers' sensitivities to prices, promotions, brands, and other marketing mix elements is fundamental to a wide swath of marketing problems. An important but understudied aspect of this problem is the dynamic nature of these sensitivities, which change over time and vary across individuals. Prior work has developed methods for capturing such dynamic heterogeneity within product categories, but neglected the possibility of \textit{correlated} dynamics \textit{across} categories. In this work, we introduce a framework to capture such correlated dynamics using a hierarchical dynamic factor model, where individual preference parameters are influenced by common cross-category dynamic latent factors, estimated through Bayesian nonparametric Gaussian processes. We apply our model to grocery purchase data, and find that a surprising degree of dynamic heterogeneity can be accounted for by only a few global trends. We also characterize the patterns in how consumers' sensitivities evolve across categories. Managerially, the proposed framework not only enhances predictive accuracy by leveraging cross-category data, but enables more precise estimation of quantities of interest, like price elasticity.  
\end{abstract}
\endgroup

\doublespacing
\newpage

\ifjmr
\else
\section{Introduction}
\fi

We know that customers' sensitivities to the marketing mix are not static. Dynamics in sensitivities can stem from global, individual-specific, or even category-level forces. For instance, fluctuations in the macroeconomic environment may drive many consumers simultaneously toward budget brands, or increase their sensitivity to prices and promotions. A consumer who has recently graduated from college and started a job may become less price sensitive or less promotion sensitive, as a result of an increase in her available budget, or may switch brands as she learns about them in a new shopping environment. Advertising and the media may drive trends, like increasing preferences for organic foods or for a specific brand in a specific category. These changes can have important implications for marketers and retailers, as understanding these shifts in preferences can help marketers optimize their targeting strategies and their allocation of marketing mix variables. 

While some forms of dynamics can be rationalized using standard economic models of utility maximization under constraint or consumer learning, it is challenging to account for such dynamics in the indirect utility models commonly used by marketers to optimize the marketing mix. This challenge has led to methodological work on dynamic heterogeneity specifications that allow for modeling individual-level dynamics in preference parameters \citep{dew2020modeling}. However, existing models often overlook a crucial factor: the correlation of consumer sensitivity dynamics across different categories. As illustrated by the simple examples above, changes in consumers' budgets or in the market as a whole may impact many categories at once, inducing correlations in the dynamics of choice parameters. Capturing such correlations can be extremely valuable to marketers: in many shopping contexts, including in both traditional consumer packaged goods and modern online settings, marketers may observe an abundance of information about a customer in one product category, but no or relatively little information in another. Understanding to what degree a shift in a consumer's preferences and sensitivities to marketing mix variables holds across product categories can help marketers leverage data in data-rich categories to make predictions about data-scarce categories. Moreover, if correlations in preferences can be accounted for by market-level trends, exploring and extrapolating those trends can be useful for market research and retail planning. 

Thus, in this paper, we develop a method for capturing correlated dynamics in individual-level marketing sensitivities, which we apply in the context of indirect utility models of brand choice. Specifically, we model individual-level parameters as weighted combinations of dynamic factors. These weights are individual-specific and hierarchically estimated, and may be correlated across different preference parameters. We call this specification the Hierarchical Dynamic Factor, or HDF, model. Echoing the previous examples, the dynamic factors can be global, meaning shared across individuals and product categories; category-specific, meaning shared across individuals, but not across product categories; or individual-specific, meaning shared across product categories, but not across individuals. Drawing from the semiparametric factor model literature in machine learning \citep{teh2005semiparametric}, and dynamic heterogeneity research in marketing \citep{dew2020modeling}, we use Bayesian nonparametric Gaussian processes (GPs) to model the latent factors. The result is a novel variant of a multi-output GP model \citep{alvarez2012kernels}, a class of models which has seen no prior applications in marketing. The resulting framework not only allows for measuring the extent of correlated dynamics across parameters and product categories, but for leveraging those correlations to improve both predictions of future choices and inferences of important quantities like price elasticity. 

We apply HDF to individual-level grocery purchase data from January 2001 through December 2012, focusing on seven common product categories. To understand the benefits and limitations of HDF, we compare its performance and insights to the Gaussian Process Dynamic Heterogeneity (GPDH) model proposed in \cite{dew2020modeling}, which, to our knowledge, is the only extant method of modeling dynamic heterogeneity in a fully flexible fashion. We find that a simple version of HDF with four global latent factors, and no category or individual-specific factors, not only yields the best predictive performance, but can explain the vast majority of the dynamic heterogeneity uncovered by GPDH, suggesting that dynamics in sensitivities can be primarily attributed to a few underlying global trends. Substantively, we find some evidence that, during the period studied, these trends relate to macroeconomic conditions, notably the Great Recession of 2008. HDF also uncovers surprising correlations in the dynamics of preference parameters, particularly across categories, that shed light on which categories and brands are responding to these global trends. Most importantly, the parsimonious factor structure of HDF, together with its ability to pool statistical information across product categories, yields better predictions for customers with sparse data, and more statistically efficient estimates of dynamics in important quantities of interest, like price elasticity. This increase in efficiency enables more precise insights as to how individuals' preferences may be changing over time. 

The remainder of this paper is organized as follows. We begin by reviewing related literature, focusing on the substantive issues of estimating cross-category effects, sensitivity dynamics, and consumer trends. We then develop our HDF model, and highlight in more detail its connections to extant specifications. Next, we showcase HDF in an application to grocery purchasing. We conclude with a summary and directions for future research.

\section{Literature} \label{sec:lit}

Our paper is informed by three streams of literature in marketing: (1) understanding and decomposing cross-category preferences; (2) estimating dynamics in sensitivities to the marketing mix, and in particular, dynamic heterogeneity; and (3) identifying latent trends from consumer data. 

\subsection{Cross-Category Preferences}

The cross-category nature of preferences has been extensively examined in marketing, beginning with \cite{ainslie1998similarities}, who investigate whether households have similar sensitivities to marketing mix variables across categories. They use a variance components decomposition approach to model households' sensitivities to these variables, and find large cross-category correlations in sensitivities to price (0.32) and to feature/promotion (0.58). Building on these findings, \cite{seetharaman1999investigating} show that households' state dependence behaviors also show large correlation across categories, and \cite{singh2005modeling} and \cite{hansen2006understanding} show the same for many other traits, including brand names, health claims, and preference for the store brand. \cite{iyengar2003leveraging} also find strong parameter correlations across product categories, but find that leveraging such correlations does not yield substantial improvements in targeting strategies. As a whole, this stream of work suggests that customers behave similarly across categories in many respects. 

Other research has examined how to model multi-category purchasing, and the complementarity or substitutability of different product categories. Early work in this vein includes \cite{manchanda1999shopping}, who model multicategory purchase incidence and examine the cross-effects that changing prices in one category may have on purchasing in another category. Building on that, \cite{mehta2007investigating} proposes a more general model that includes both category purchase incidence and brand choice decisions, and \cite{song2007discretecontinuous} integrate purchase quantity decisions into the framework. \cite{lee2013direct} further build on the previous work and allow the cross-effects between categories to be asymmetric. While these papers do not explicitly model the linkages between preferences in disparate product categories, they suggest that consumers make decisions at the basket-level, not the category-level, suggesting that changes in the decision-making process should simultaneously affect multiple categories at once.

Our research is aligned more closely with the first stream of research: rather than positing a single model of cross-category purchasing, we model brand choice within categories separately, allowing those decisions to be linked by correlations in the customer-level parameters governing those choices. Our key contribution is exploring customers' preference \textit{dynamics}; that is, how \textit{changes} in preferences in one category are indicative of \textit{changes} in another. All prior work has looked at cross-category correlations in static parameters. Methodologically, our work introduces a dynamic analogue to the cross-category factor model of \cite{singh2005modeling}: by identifying a dynamic factor regime that explains variation in purchasing, we can attribute preference variation to global, category, or individual-level trends.

\subsection{Preference Dynamics}

Our work also contributes to a literature investigating how consumers' sensitivities to marketing variables evolve over time. An early contribution in this area is \cite{mela1997long}, who show that consumers' sensitivities not only evolve, but that firm actions can drive such changes. More recent work has also shown that the macroeconomic environment is another driver of dynamics. For instance, \cite{kamakura2012economic}, show that budget allocations across different product categories change during economic downturns and expansions, with expenditures on positional products --- that is, products used to signal societal position --- decreasing during economic downturns. \cite{gordon2013does} investigate the relationship between price elasticity and the business cycle, finding that, on average, customers are more price sensitive during economic downturns, with a few notable exceptions for small share-of-wallet categories. Other recent papers have examined the effect of a specific type of economic downturn --- recessions, and more specifically the 2008 Great Recession --- on consumer behavior. Contributions in this stream include \cite{dube2018income}, who document the causal, negative effects of wealth and income on private label demand, focusing on the period of the Great Recession. Interestingly, their work also documents a positive trend in private label demand, predating 2008. \cite{nevo2019elasticity} also examine the effect of the Great Recession on demand, finding higher rates of purchasing on sale, in larger quantities, of more generic products, using more coupons, and from more discount stores. Finally, \cite{dew2020modeling} show a substantial effect of the 2008 Great Recession on preferences, most notably on price elasticity.  Together, this literature suggests that sensitivities to the marketing mix do change, for a variety of reasons, and that capturing such changes is critical for marketing mix decisions. 

Our work provides a new method for capturing such sensitivity dynamics, which can allow for more precise estimation of individual-level effects. As such, we build on a long stream of research developing ways of estimating time-varying preference parameters from choice data, including \cite{kim2005modeling}, \cite{liechty2005dynamic}, \cite{sriram2006effects}, \cite{lachaab2006modeling}, and \cite{guhl2018estimating}. While these papers all document evidence of changing sensitivities over time, their modeling assumptions are also restrictive: they only model dynamics at the population level, and capture heterogeneity in a static way, as individual-level (static) effects around a shared trend. To allow for the \textit{heterogeneous} evolution of such parameters, \cite{dew2020modeling} introduce the idea of dynamic heterogeneity, or a continuously evolving distribution of unobserved heterogeneity. They model dynamic heterogeneity using Gaussian processes, in what they term the Gaussian Process Dynamic Heterogeneity (GPDH) specification, which allows individual-level parameters to differ from the population mean parameter at different time periods in a parsimonious, hierarchical Bayesian fashion. They show that not only are there substantial preference dynamics at the individual-level, but that ignoring dynamic heterogeneity can bias parameter and elasticity estimates. While the GPDH model relaxes one restrictive assumption in modeling preference dynamics, it retains others. Most notably, the GPDH model considers only one category at a time, and ignores potential correlations in how preferences evolve. 

Building on this body of work, we develop a cross-category, individual-level, Bayesian nonparametric model that captures the dynamics of customers' preference parameters in a given category, as well as the cross-category nature of the dynamics in preferences, through a dynamic latent factor structure. Through this model, we allow information learned about how customers' preference parameters have evolved in one category to be transferred across categories, and inform predictions of preference evolution in another category. Such information sharing allows us to forecast customers' brand choices more accurately, and produces more reliable estimates of quantities of interest like price elasticity. 

\subsection{Identifying Trends}

Finally, our work also relates to a small literature on identifying market trends from data, and on dynamic factor models more generally. Closest to ours is the work of \cite{du2012quantitative}, who develop a structural dynamic factor model for isolating trends from time series data. In their application, they apply their model to Google search data, to understand long term trends in vehicle shopping. Our work builds on this dynamic factor idea, but in a choice context, and where the object of interest is individual-level preference parameters, rather than observed time series.\footnote{We discuss the specific methodological distinctions between our model and theirs, as well as a host of other specifications from machine learning, in the following section.} Other important contributions in this literature include \cite{schweidel2014listening}  and \cite{zhong2020capturing}, who propose ways of measuring dynamic brand sentiment and other perceptions from social media data.

\section{Model} \label{sec:model}

We now present our Hierarchical Dynamic Factor (HDF) model for capturing correlated dynamics in preferences. We begin by describing the general model specification. The heart of the model is a set of dynamic factors, which we model with Bayesian nonparametric Gaussian processes (GPs), following the literature on dynamic heterogeneity \citep{dew2020modeling}. Because GPs are relatively uncommon in marketing, following the general specification, we review GPs and describe the details of our GP-based dynamic factor specification. We then give the full set of priors for each part of the model, which complete the specification. Finally, we describe how HDF relates to several extant methods, in marketing, statistics, and machine learning.

\subsection{Hierarchical Dynamic Factor (HDF) Model}

Following the standard random (indirect) utility specification, we assume that the utility that customer \(i\) gets from purchasing product \(j\) in category \(c\) at time \(t\) is:
\begin{equation} \label{eq:logit}
u_{icjt} =\sum_{p=1}^{P_c} \beta_{icpt} \times x_{pcjt}+\epsilon_{icjt}, 
\end{equation}
where \(\epsilon_{icjt}\) is IID extreme value, leading to the standard logit choice probabilities, and where $p$ indexes parameters. $P_c$ is the number of predictors of brand choice in category $c$. In our empirical application, \(x_{pcjt}\) includes the marketing mix variables price and promotion, as well as brand level dummy variables, and thus \(\beta_{icpt}\) measures customer \(i\)'s sensitivity to variable \(p\) in category \(c\) at time \(t\). For identification purposes, we normalize the brand dummy variable of one brand to 0 across all time periods for all individuals, such that all other brand intercepts are interpreted relative to this brand. For concision, going forward, we will concatenate the $c$ and $p$ indices to a single index, \(s = (c,p)\), such that $s = 1, \ldots, S$, where $S = \sum_c P_c$, and \(\beta_{ist}\) is the $s$th sensitivity of interest. 

To parsimoniously model \(\beta_{ist}\) in a way that not only captures individual-specific patterns of time variation, but also allows for correlations over $s$, we first make a notational pivot, and consider $\beta_{is}$ as a function of $t$, i.e., \(\beta_{ist} = \beta_{is}(t)\). This notational pivot makes the connection to Gaussian processes clearer, as we discuss in the next section. Then, we model $\beta_{is}(t)$ through a set of dynamic factors. We begin by describing the simplest case, where those factors are shared across all individuals and categories, before describing extensions with category- and individual-specific factors.

\paragraph{HDF With Global Factors} The heart of our proposed model is to specify $\beta_{is}(t)$ as a weighted sum of $L$ latent factors, where the weights on those latent factors are individual and parameter-specific:
\begin{equation} \label{eq:hdf_beta}
  \beta_{is}(t) = \alpha_{is} + \sum_{\ell=1}^{L} \omega_{is\ell} \times \theta_\ell(t).
\end{equation}
Here, $\alpha_{is}$ captures the time-invariant \textit{level} of parameter $s$ for person $i$, and the \(\theta_\ell(t)\) are latent functions capturing \textit{dynamics}, modeled as Gaussian processes, as we describe subsequently. Note that, here, $\theta_\ell(t)$ is shared across all customers and parameters. Recalling that $s = (c,p)$, this implies that these latent factors are also shared across all categories. In other words, they are \textit{global} latent factors. 

The links between the factors and the individual-level preference parameters are the weights, $\omega_{is\ell}$. We constrain these weights to be positive, to aid in interpretability and identification, and model them hierarchically. We do so by first denoting an unconstrained variable, $\tilde{\omega}_{is\ell}$, and then defining $\omega_{is\ell} = \log(1+\exp(\tilde{\omega}_{is\ell}))$. This link function is sometimes called the ``softplus,'' and is a monotonic, positive function \citep{dugas2000incorporating}.\footnote{The softplus is used instead of $\exp(\cdot)$ for increased numeric stability.} Then, we model $\tilde{\omega}_{is\ell}$ as:
\begin{equation} \label{eq:weight_vector_prior}
  \boldsymbol{\tilde{\omega}}_{il} = (\tilde{\omega}_{i1l}, \ldots, \tilde{\omega}_{isl}) \sim \mathcal{N}(0,\Sigma_{\omega}), 
\end{equation}
where \(\Sigma_{\omega}\) is a \(K \times K\) covariance matrix, that captures correlations in the weights across parameters.\footnote{$\Sigma_\omega$ might be more correctly denoted as $\Sigma_{\tilde{\omega}}$, as it defines a covariance over $\boldsymbol{\tilde{\omega}}_{il}$; we favor the slightly simpler notation.} The combination of the hierarchical weights and the dynamic latent factors gives us the Hierarchical Dynamic Factor (HDF) model. 

The HDF model structure has two important features. First, while the latent factors $\theta_\ell(t)$ are assumed to be independent, their convolution through the weights results in a set of correlated functions, comprising the dynamic sensitivities, \(\beta_{is}(t)\). In other words, because each $\beta_{is}(t)$ is modeled using the same set of latent factors, they will be correlated. Second, the nature of this correlation is determined, in part, by $\Sigma_\omega$. Intuitively, $\Sigma_\omega$ is an $S \times S$ matrix, where entry $(s,s')$ captures the covariance between the weights of parameters $s$ and $s'$ on factor $\ell$. If this covariance is high, it means these parameters tend to place weight on the same factors. Understanding these covariances is one of the important outputs of the model, as they shed light on which sensitivities in which categories have correlated dynamics.

Beyond inducing correlations in parameters, the HDF has other advantages. Foremost, the use of a few common latent factors in modeling dynamics provides a source of dimensionality reduction, avoiding the need to model each \(\beta_{is}(t)\) as a separate function. This structure pools information across different customers, and across different points in time. The weights also provide a source of pooling: since the weight vector \(\boldsymbol{\omega}_{il}\) is modeled hierarchically, with a full covariance matrix, it allows for information sharing across preference parameters, including those of different product categories. Moreover, since the covariance is shared across people, it provides yet another source of pooling. Together, these mechanisms allow us to turn the complex problem of estimating \(\beta_{is}(t)\) into a simpler problem, with considerable information sharing across units of analysis. In turn, this yields increased precision in model estimates, especially for customers with limited or sparse data, as we will demonstrate empirically in later sections. From a substantive perspective, modeling the dynamics in preferences through a set of common latent factors has another advantage: if there are common temporal shocks to many consumers' preference parameters, these may be naturally captured in one or several of the latent factors, and consumers' individual weights on those factors may thus indicate to what degree they were affected by those temporal shocks. 

\paragraph{Adding Individual and Category Factors} The model above consists of only global factors. As we will see in the empirical application, these global factors alone can explain a substantial portion of the dynamic heterogeneity observed in our data. That being said, it is possible to extend the model to consider other types of dynamic factors. We consider two other types of factors: category-specific and individual-specific. Category-specific factors are dynamic factors that drive parametric evolution within a single product category, across individuals. Likewise, individual-specific factors drive parametric evolution within a single person, across categories. 

Extending the HDF to capture these effects is straightforward. First, expand our notation. We now have three types of latent factors: global factors, $\theta^\mathrm{Glob}_{\ell_G}(t)$; category factors, $\theta^\mathrm{Cat}_{c \ell_c}(t)$; and individual factors, $\theta^\mathrm{Ind}_{i \ell_i}$. The different subscripts, $\ell_G$, $\ell_c$ and $\ell_i$ indicate that the number of latent factors need not be the same within each group. We will denote the total number of latent factors globally as $L_G$, for category $c$ as $L_c$, and for person $i$ as $L_i$ respectively.\footnote{In practice, allowing there to be variable number of latent factors, especially for individual-level factors, may be challenging to implement. In our empirical applications, we will always assume the same number of latent factors per category or per individual.} To incorporate these factors into the model, we return to the original, full subscript notation:
\begin{equation}
  \beta_{icp}(t) = \alpha_{icp} + \sum_{\ell_G=1}^{L_G} \omega^\mathrm{Glob}_{icp\ell_G} \times \theta^\mathrm{Glob}_{\ell_G}(t) + \sum_{\ell_c=1}^{L_c} \omega^\mathrm{Cat}_{icp\ell_c} \times \theta^\mathrm{Cat}_{c \ell_c}(t) + \sum_{\ell_i=1}^{L_i} \omega^\mathrm{Ind}_{icp\ell_i} \times \theta^\mathrm{Ind}_{i \ell_i}(t).
\end{equation}
While the notation is somewhat complex, the idea is simple: category-specific factors only influence parameters from their respective category, and likewise for individual-specific factors. Otherwise, the model is the same as before. 

Having defined the general form of the HDF and its different types of factors, we now turn to three remaining aspects of model specification: (1) how to actually model the latent factors, $\theta_\ell(t)$; (2) how to determine an optimal number of latent factors, which directly affects the flexibility of the model; and (3) how to specify the priors for the remaining model components, most importantly the covariance, $\Sigma_\omega$. For simplicity, in the remainder of this section, we will assume a model with only global factors, though the ideas apply analogously to the more complex variants of HDF described above.

\subsection{Gaussian Process Latent Factors}

To model the latent factors, we use Bayesian nonparametric Gaussian processes. Gaussian processes, or GPs, provide a way of putting a prior over an unknown function of interest. GPs have many desirable properties for modeling parametric dynamics, and, in this case, latent factors, and have seen increasing use in marketing for these purposes \citep[e.g.,][]{dew2020modeling, dew2023detecting}. We first review the basics of GPs, before describing the specific GP priors used in HDF. 

\paragraph{GP Primer} In brief, a Gaussian process (GP) is a stochastic process \(f(.)\) defined on an input space, which, in our case, is time \(t \in R^+\). While GPs can be defined over any dimensional input space, we will focus throughout on the simple case where the input is a scalar. A GP is defined by a mean function \(m(t)\) and a covariance function \(k(t,t')\), such that for a fixed set of inputs \(\boldsymbol{t} = \{t_1, t_2, \ldots, t_T\}\),
\[ 
f(\boldsymbol{t}) \sim \mathcal{N}(m(\boldsymbol{t}),K(\boldsymbol{t})),
\]
where \(m(\boldsymbol{t})\) is the mean function evaluated at all inputs, a \(T \times 1\) vector, and \(K(\boldsymbol{t})\) is a \(T \times T\) matrix formed by evaluating the covariance function \(k(t,t')\) pairwise at all of the inputs. In short, the mean function specifies the prior expectation of the value of the process for each input $t$, and the covariance function specifies how correlated the process is across pairs of inputs, \(t\) and \(t'\). Since a GP defines a probability distribution over outputs, given any inputs, it serves as a natural prior over functions \citep{rasmussen2006gaussian}. In this capacity, is typically denoted \(f(t) \sim \mathcal{GP}(m(t), k(t,t'))\).

In most GP applications, mean functions play a limited role, and are usually assumed to be constant, allowing the covariance function to capture the properties of the functions generated by the GP priors. Different covariance functions, also called kernels, can capture different broad features of the functions being modeled. A valid kernel is a function \(k: \mathbb{R}^2 \to \mathbb{R}\) (again, for scalar inputs), such that the covariance matrix \(K(\boldsymbol{t})\) generated by the kernel function is positive definite for any set of inputs \(\boldsymbol{t}\). Many kernels have been proposed in the GP literature, but perhaps the most widely used is the Mat\'ern kernel. They are used to model stationary processes (i.e., processes that, in the long run, revert to their mean) by means of three parameters, which govern three important properties of the resulting functions: amplitude, smoothness, and differentiability. Mathematically, the Mat\'ern kernel is given by: 
\[
k(t,t';\eta,\kappa,\nu) = \eta^2 \frac{2^{1-\nu}}{\Gamma(\nu)} \, \left[\rho \lvert t-t' \rvert\right]^{\nu} \, \mathit{K}_{\nu}(\rho \lvert t-t' \rvert),
\]
where $K_\nu(\cdot)$ is the modified Bessel function of the second kind, and where $\eta$ controls the amplitude, $\rho$ controls the (inverse) smoothness, and $\nu$ controls the level of differentiability.\footnote{The parameterization here is called the inverse lengthscale parameterization of the Mat\'ern kernel. In many other applications, the Mat\'ern is parameterized as $k(t,t';\eta,\kappa,\nu) = \eta^2 \frac{2^{1-\nu}}{\Gamma(\nu)}\left(\frac{\sqrt{2\nu}|t-t'|}{\kappa}\right)^\nu K_\nu\left(\frac{\sqrt{2\nu}|t-t'|}{\kappa}\right)$, and the parameter $\kappa$ is called the lengthscale. We prefer the inverse lengthscale parameterization, $\rho = \sqrt{2\nu}/\kappa$, for ease of specifying its prior. This specification is also the same as in \cite{dew2020modeling}.} This intimidating functional form dramatically simplifies when $\nu$ is chosen to be a half-integer, most commonly $1/2$, $3/2$, or $5/2$. For instance, when $\nu = 3/2$ (as in our model), the resulting kernel is given by:
\[
k(t,t';\eta,\kappa) = \eta^2 (1+\kappa\lvert t-t' \rvert)\exp(-\kappa\lvert t-t' \rvert)
\]
In this form, the roles of $\eta$ and $\rho$ are more apparent: if $t = t'$, the kernel reduces to just $\eta^2$. Hence, $\eta^2$ controls the variability of the outputs. Otherwise, the magnitude of the kernel is determined by $\rho^2 |t - t'|$. In this way, $\rho$ controls how quickly the covariance degrades as $t$ and $t'$ become farther apart --- the smaller $\rho$, the slower the degradation, and thus the smoother the function. For this reason, $\rho$ is called the \textit{inverse lengthscale} parameter. More generally, the parameters of the kernel, like $\eta$ and $\rho$, are typically called the \textit{kernel hyperparameters}. 

\paragraph{GP Priors in the HDF Model}

In our specification, we model each $\theta_\ell(t)$ as a GP, with a constant mean and a Mat\'ern-3/2 kernel (i.e., a Mat\'ern with $\nu = 3/2$), with amplitude $\eta$ fixed to 1:
\begin{equation}
  \theta_\ell(t) \sim \mathcal{GP}(m_\ell, k(t,t'; 1, \rho_\ell)).
\end{equation}
We restrict the amplitude to 1 for identification, since we do not restrict the values of the weight terms $\omega$. Including a constant mean $m_\ell$ versus a zero mean helps with out-of-sample predictions, but is otherwise not essential. For the kernel, we choose the Mat\'ern-3/2 kernel to be consistent with \cite{dew2020modeling}; in our empirical application, we experimented with different types of Mat\'ern kernels (including the popular squared exponential kernel, which is the limiting case of the Mat\'ern as $\nu \rightarrow \infty$), and found the results robust to this choice. 

\begin{figure}
  \centering
  \includegraphics*[width=0.6\textwidth]{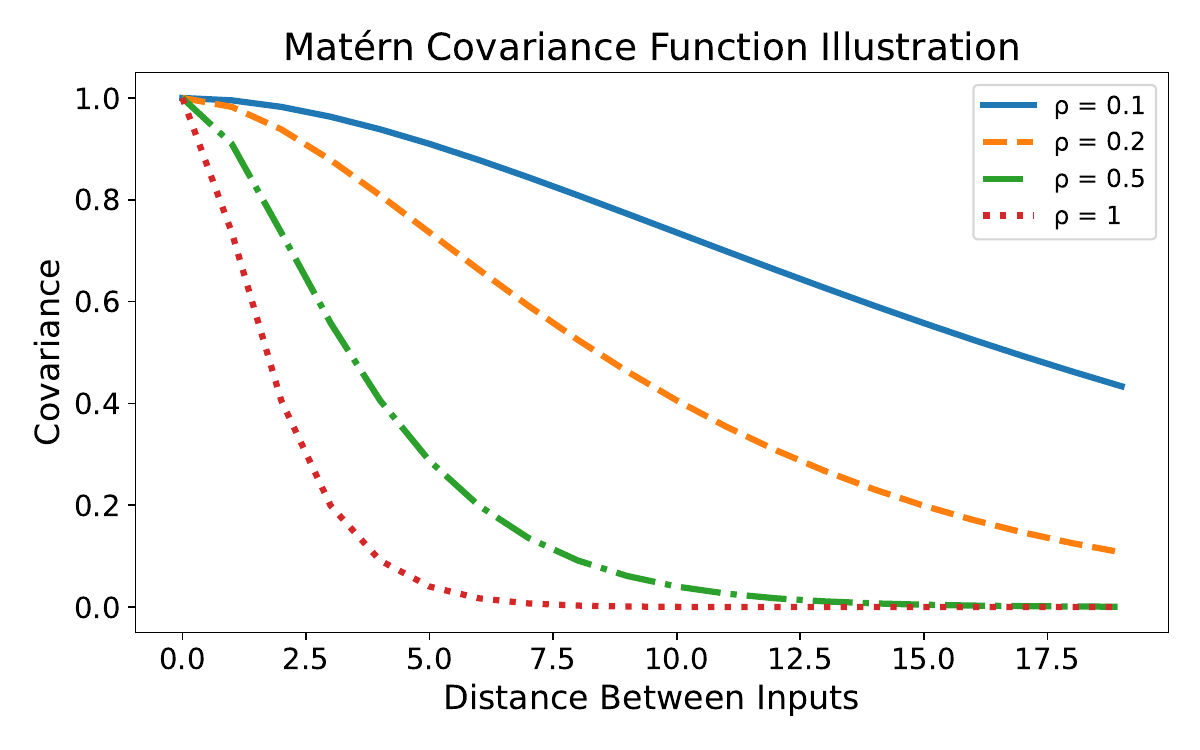}
  \caption{Effect of $\rho$ on Smoothness: Each curve shows how the covariance between GP function values decays as a function of $|t-t'|$ for a different value of $\rho$.}
  \label{fig:matern_rho_illustration}
\end{figure}

As with all parameters in the model, we estimate the kernel hyperparameter, $\rho_\ell$, in a fully Bayesian fashion. Thus, we must assign it a prior. This prior is important: $\rho$ controls the degree of intertemporal regularization via the smoothness of the resulting factors. To illustrate this regularization, in \Cref{fig:matern_rho_illustration}, we show how the covariance of the function values associated with inputs $t$ and $t'$ decays as a function of $|t - t'|$, under different values of $\rho$. As $\rho$ goes down, this covariance decays more slowly, leading to smoother functions. Thus, setting the prior for $\rho$ is equivalent to setting the level of regularization for the dynamics in the model. To this end, we choose a fairly informative Weibull prior:
\begin{equation}
  \rho_\ell \sim \mathrm{Weibull}(0.1, 1).
\end{equation}
This choice is motivated by several factors: first, the Weibull distribution emerges as the natural prior for the inverse lengthscale when the goal is to penalize the complexity of the resulting functions, and when the amplitude is fixed \citep{fuglstad2019constructing}.\footnote{In that sense, the Weibull is equivalent to the penalize complexity prior used in \cite{dew2020modeling}.} Second, with certain parameters, the Weibull places its probability mass toward zero, which is equivalent to putting high probability on smooth latent factors. The specific parameters we use for the Weibull --- shape 1, scale 0.1 --- imply that 90\% of the prior mass is between [0.005, 0.3], leading to a priori smooth factors.

\subsection{Determining the Number of Latent Factors} \label{sec:num-latent-factors}

\begin{figure}
  \centering 
  \includegraphics[width=0.49\linewidth]{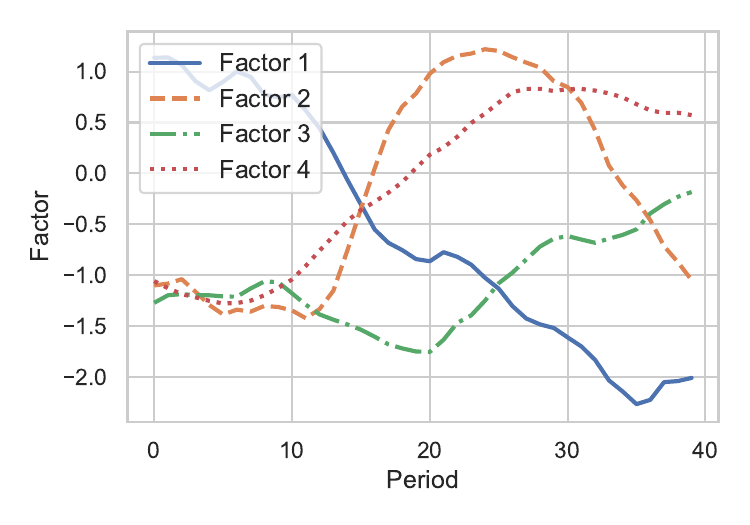}
  \includegraphics[width=0.49\linewidth]{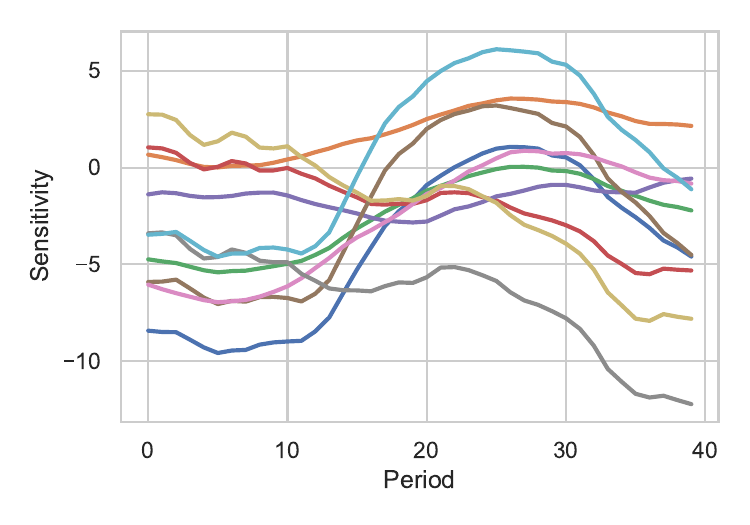}
  \caption{Simulation of four latent factors and the preference parameters generated by them. Latent factors are plotted on the left, and each line type represents a different latent factor. Preference parameters are plotted on the right, and each color represents a different preference parameter.}\label{fig:plot-of-simulated-draws}
\end{figure}

A second crucial part of HDF model specification is setting the number of latent factors. One interpretation of the latent factors is as basis functions: as described in \Cref{eq:hdf_beta}, $\beta_{is}(t)$ is comprised of an intercept plus a linear combination of nonlinear functions. Hence, with enough latent factors, any pattern of temporal variation in $\beta_{is}(t)$ should be able to be captured. In this way, the HDF is quite flexible. In fact, even with relatively few latent factors, the HDF can capture a wide variety of individual-level parameter trajectories. We illustrate this point using a simulation with four latent factors, in \Cref{fig:plot-of-simulated-draws}. In the left panel, we show four draws from GPs, representing four latent factors. At right, we show 10 random linear combinations of these factors, with weights drawn exactly as in the model. While there are obviously \textit{some} similarities among some of the trajectories in the right panel -- for instance, there are clearly several that load highly onto the hill-like Factor 2 (in the left panel) -- they are notably variable. 

The risk of adding more factors is potential overfitting. One important safeguard against overfitting is the prior on the inverse lengthscale: the Weibull prior sets a strong prior toward very smooth functions, such that, absent information, latent factors will be very smooth, and thus not contribute significantly to the dynamics.\footnote{This idea is similar to the way that automatic relevance determination (ARD) models work \citep{mackay1994bayesian}.} In our empirical application, we test a number of latent factors, and compare predictive performance. We find that forecasting performance does eventually degrade with more factors, but not rapidly, suggesting that the model is reasonably robust to this choice. Even so, care should be still be taken to pick a good, minimal number. 

\subsection{Other Priors}

To complete the HDF specification, we need priors for the remaining model parameters, including $\Sigma_\omega$. We now briefly discuss these priors.

\paragraph{Prior for $\Sigma_\omega$} 
We model $\Sigma_\omega$ as: 
\begin{equation} \label{eq:corr_mat_weights}
  \Sigma_\omega = \mathrm{diag}(\boldsymbol{\tau}) \, \Lambda_\omega \, \mathrm{diag}(\boldsymbol{\tau}),
\end{equation} 
where $\mathrm{diag}(\boldsymbol{\tau})$ refers to the diagonal matrix with the vector $\boldsymbol{\tau}$ along the diagonal. $\boldsymbol{\tau}$ captures the scale of variation of preference sensitivities across customers, while $\Lambda_\omega$ is a correlation matrix that captures the correlations in the dynamics between preference sensitivities. We use an uninformative prior for $\boldsymbol{\tau}$, assigning each of elements a $\mathrm{HalfNormal}(5)$ prior (i.e., the positive part of a normal distribution, with variance 5). For $\Lambda_\omega$, we use an $\mathrm{LKJ}(2)$ prior. The LKJ distribution is a distribution over correlation matrices that has a single shape parameter \citep{lewandowski2009generating}. By setting that shape parameter to 2, we are setting a weakly informative prior favoring the identity matrix.

\paragraph{Priors for $\alpha_{is}$} For the HDF level parameters, we use a multivariate normal prior, such that:
\begin{equation}
  \boldsymbol{\alpha}_i = (\alpha_{i1}, \ldots, \alpha_{iS}) \sim \mathcal{N}(\boldsymbol{\mu}_\alpha, \Sigma_\alpha).
\end{equation}
$\boldsymbol{\mu}_\alpha$ is given a component-wise $\mathcal{N}(0,5)$ prior, while $\Sigma_\alpha$ is treated similarly to $\Sigma_\omega$. Notably, if all of the weights on the latent factors are zero, the model collapses to $\beta_{is}(t) = \alpha_{is}$, a mixed logit model with a full covariance matrix. This model is one of our benchmarks later. 

\subsection{Estimation}

We estimate the model in a fully Bayesian fashion, using Hamiltonian Monte Carlo \citep{betancourt2018conceptual} via the No-U-Turn sampler (NUTS). Specifically, we implement the model in the NumPyro probabilistic programming language, which allows us to leverage recently introduced GPU-enabled accelerated computation for scalability \citep{phan2019composable}.\footnote{The GPU implementation yields considerable gains in performance: using the CPU-based Stan implementation of the GPDH dynamic heterogeneity specification from \cite{dew2020modeling}, estimating the model on data from a single category for ~1,000 customers and ~25,000 purchases takes days. In contrast, our full HDF model, run for 1,000 sampling iterations, using data from over 3,000 customers across 7 product categories for a total of 1.6M purchase observations, takes 3 hours. For practical use, fewer iterations can be used, further reducing the time to estimate the model.} A full implementation can be found at [GITHUB LINK REDACTED FOR PEER REVIEW]. 

\subsection{Comparisons to Existing Methods} \label{sec:model-comparisons}

Methodologically, the HDF model sits at the intersection of three types of models: multi-output GPs, models of dynamic latent factors, and models of dynamic heterogeneity. In this section, we highlight the links and distinctions between the present work and these other approaches.

\paragraph{Multi-output GPs} A multi-output Gaussian process (or MOGP) is, as the name suggests, a generalization of the standard GP to work with multiple outputs \citep{alvarez2012kernels}. That is, rather than modeling scalar-valued functions, as in an ordinary GP, MOGPs model vector-valued functions. The HDF model can be viewed as a specific type of MOGP: each $\beta_{is}(t)$ is modeled as a sum of GPs, and just as sums of Gaussians are Gaussian, sums of GPs are GPs. Thus, HDF is a MOGP model, where the outputs are consumers' preference trajectories across categories. 

While MOGPs, to our knowledge, have never been used in marketing applications, they have a long history in statistics and machine learning \citep{alvarez2012kernels}. In the MOGP literature, the idea of linearly combining independent factors to learn correlated processes is called the linear model of coregionalization (LMC) \citep{alvarez2012kernels}, an idea originally inspired by the cokriging models of geostatistics \citep{journel1976mining, wackernagel2013multivariate}. In machine learning, \cite{teh2005semiparametric} propose a similar idea to the LMC in their semiparametric latent factor model (SLFM). 

Our specification for correlated dynamics is inspired by these models, but with two important distinctions. The first key difference is scale: while most geostatistical and machine applications model a handful of related outcomes, our model focuses on modeling many consumers, each of whom has many potentially correlated sensitivities. To handle this scale, we introduce the hierarchical structure on the weights, that allows for correlated patterns within individuals, and sharing of information across individuals. The second distinction is that, in both in geostatistics and in machine learning, models like the LMC and SLFM are almost exclusively used in regression or classification tasks, where the outcomes of interest are directly observed, and modeled (up to some noise) by the linear combination of GPs. In our application, the task is much more difficult: our ``outcome'' of interest is, in fact, a model coefficient capturing consumer sensitivities, which are then combined with observed variables (i.e., the marketing mix) to model utilities which determine choice. The jump from modeling the regression function directly to modeling latent quantities like model parameters is non-trivial. 

\paragraph{Dynamic Factor Models}

As discussed previously, our model is also similar to the structural dynamic factor model of \cite{du2012quantitative}. In their work, a time series specification forms the basis of a set of latent factors, which in turn, drive observed trends in Google search volume. Our HDF model is similar, insofar as a common set of dynamic factors drives a related set of outcomes; however, those outcomes are the parameters underpinning a choice process, not an observed time series. Nonetheless, our work can be viewed as an extension of theirs to this limited dependent variable setting. Another key difference is how we model the latent factors: rather than using a time series specification, we use GPs. Our choice of GPs is motivated by their good performance both in terms of modeling dynamic heterogeneity, and in the multi-output settings discussed above. That being said, the GPs in our method could be replaced by other dynamic models, including, for instance, state space and time series specifications; exploring alternative dynamic factor specifications may be an interesting area of future research.\footnote{In developing our model, we tested several simple state space specifications for $\theta_\ell(t)$, including a simple autoregressive process. We found the model performed worse than the GP version, but do not include this model as a benchmark.}

\paragraph{Dynamic Heterogeneity}

Perhaps closest to the current work is the Gaussian Process Dynamic Heterogeneity specification of \cite{dew2020modeling}, on which the present work builds. In their GP Dynamic Heterogeneity (GPDH) specification, consumers' sensitivities to marketing mix variables are modeled as:
\begin{equation}
  \beta_{is}(t) \sim \mathcal{GP}(\mu_s(t), k_s(t,t'; \phi)),
\end{equation}
where the $\mu_s(t)$ is a mean model, estimated from the data, and the $k_s()$ is, as noted previously, the Mat\'ern-3/2 kernel. They explore several different mean models, including GPs and state space models.

Our work provides an alternative method for modeling such dynamic heterogeneity. While the GPDH model is more flexible than HDF, the HDF model is simpler, as all dynamics stem from just a few factors. This simplicity can mitigate overfitting concerns, provide additional insight into the drivers of dynamic heterogeneity via the latent factors, and substantially reduce the model's computational burden. HDF also allows for correlations across preference parameters, which can improve efficiency with limited data. As the GPDH currently represents the gold standard for estimating dynamic heterogeneity, comparing these two specifications is the focus of much of our empirical analysis below.

\section{Empirical Application: Brand Choice in Groceries}
\label{sec:data}

We apply the HDF model to brand choice in the IRI consumer packaged goods panel data, from January 2001 to December 2012. The IRI data records the UPC-level transactions of customers of IRI's BehaviorScan Program.\footnote{The IRI data has been used in many papers, including \cite{gordon2013does}, which contains a detailed discussion of the data.} This period covers the Great Recession, which started from December 2007 and ended in June 2009.\footnote{\url{https://www.nber.org/research/business-cycle-dating}} We expect to detect interesting patterns of dynamics of preference sensitivities and price elasticities during this period. We restrict our analysis to the 10 stores in which IRI tracked prices throughout the 12 years. We focus on seven common product categories: coffee, paper towels, soda, toilet tissues, cold cereal, mustard, and ketchup.\footnote{These seven categories were chosen given their relatively high purchase rate across customers, but otherwise essentially at random. This provides a fair test for our model: uncovering meaningful correlated dynamics in this rather generic context provides strong evidence for the sensitivity of the model.} Since we model customer choice at the brand level, we aggregate the UPC-level choice information into brands. For each product category, we keep the transaction data of the top brands by market share. After excluding any brand whose price is missing from an IRI store for more than 100 consecutive weeks, our final panel includes 3-6 brands per category, collectively comprising 60-90\% of the market share. The top brands in each category include always include a private label brand.

In addition to brand, we include two marketing mix variables: price and promotion. To get price and promotion information for each brand, in each time period, we aggregate UPC-level price and promotion information. Promotion of a UPC is coded as a dummy variable, following the same procedure as in \cite{gordon2013does}. Specifically, promotion takes a value of 1 if a UPC's total price reduction is larger than 5\%. We form our price variable by  first normalizing the purchase price of each UPC by its volume to get a unit price. For each week in each store, we aggregate the unit price of each UPC bought, weighed by the share of the UPC compared to all UPCs of the brand in that week. We repeat the same process for the promotion variable. Thus, in our cleaned data, price and promotion of a brand varies at the store and week level. Price, promotion, and brand dummies are the explanatory variables that we use in this application. 

Following \cite{gordon2013does}, we model the time variation of preference sensitivities at the quarterly level. For this analysis, we only include customers who were active in both the starting 6 quarters and ending 6 quarters of our sample.\footnote{Such a restriction is important for ensuring that the dynamics observed represent true dynamics, not just differences in panel composition over time.} This filtering rule leaves us with 3,292 customers. Over the span of the 48 quarters, these customers made a total of 1.68 million purchases across the seven categories. Soda is the most purchased category, with an average of 190 purchases per customer over the sample period across all customers, and mustard is the least purchased category, with an average of 10.6 purchases. We include additional category-level summaries in \Cref{tab:data-summary}. 80\% of the customers purchased at least once in all 7 categories, and every customer purchased in at least 2 categories. We estimate the model with the first 44 quarters (11 years) of the data and use the last 4 quarters (i.e., year 2012) for evaluating the forecasting performance of the model.

One important aspect of any analysis of the marketing mix is endogeneity: retailers often set prices and other marketing mix variables in ways that are correlated with unobservables, leading to non-negligible correlations between the $x$ variables and the error term in \Cref{eq:logit}. There are many methods for accounting for such endogeneity, and which can be used in tandem with our HDF specification, including the control function approach of \cite{petrin2010control}. While endogeneity is a concern for making substantive claims, it is not focal to our objective, which is showing the benefits of the HDF approach to modeling dynamic heterogeneity. We do so through both predictive analyses, and by comparisons to other methods for accounting for such heterogeneity. Thus, we apply HDF to our data without any accounting for potential endogeneity. 

\ifjmr
\renewcommand{\arraystretch}{1.25}
\fi

\begin{table}[!ht]
\caption{\label{tab:data-summary} Summary of each category's purchasing data.}
\centering
\begin{adjustbox}{center}
\begin{tabular}{lrrrrrrr}
\toprule
                      & Coffee    & Paper Towels & Soda      & Toilet Tissue & Cereal    & Mustard  & Ketchup \\
\midrule
\cellcolor{gray!6}{\# Purchases}   & \cellcolor{gray!6}{120,594}   & \cellcolor{gray!6}{120,003}      & \cellcolor{gray!6}{625,283}   & \cellcolor{gray!6}{175,699}       & \cellcolor{gray!6}{531,022}   & \cellcolor{gray!6}{35,041}   & \cellcolor{gray!6}{68,531}  \\ 
\# Customers   & 2,995     & 3,185        & 3,282     & 3,241         & 3,283     & 2,934    & 3,230   \\
\cellcolor{gray!6}{\# Brands}      & \cellcolor{gray!6}{4}         & \cellcolor{gray!6}{4}            & \cellcolor{gray!6}{6}        & \cellcolor{gray!6}{5}             & \cellcolor{gray!6}{6}         & \cellcolor{gray!6}{3}        & \cellcolor{gray!6}{3}       \\
Price: Mean    & 4.32      & 2.06         & 4.44      & 0.60          & 3.21      & 2.00     & 0.96    \\
\cellcolor{gray!6}{Price: SD}      & \cellcolor{gray!6}{1.39}      & \cellcolor{gray!6}{0.61}         & \cellcolor{gray!6}{1.15}      & \cellcolor{gray!6}{0.16}          & \cellcolor{gray!6}{0.69}      & \cellcolor{gray!6}{0.57}     & \cellcolor{gray!6}{0.25}    \\
Promo: Mean    & 0.21      & 0.21         & 0.47      & 0.23          & 0.24      & 0.18     & 0.24    \\
\cellcolor{gray!6}{Promo: SD}      & \cellcolor{gray!6}{0.33}      & \cellcolor{gray!6}{0.33}         & \cellcolor{gray!6}{0.35}      & \cellcolor{gray!6}{0.33}          & \cellcolor{gray!6}{0.30}      & \cellcolor{gray!6}{0.32}     & \cellcolor{gray!6}{0.36}    \\
\bottomrule
\end{tabular}
\end{adjustbox}
\end{table}

\section{Results} \label{sec:results}

We present the results of our empirical analysis in three parts. First, we explore the predictive performance of the HDF model, both to determine the optimal specification of the model in terms of the number and types of factors, and to compare its performance to several benchmarks. Then, using the optimal model structure found in the first part, we explore what the parameter estimates imply about sensitivity dynamics and their cross-category linkages. Finally, we consider the implications of using HDF \textit{beyond} prediction, especially in terms of more efficient estimation of preference dynamics.

\subsection{Predictive Performance} \label{sec:predictive-validation}

\paragraph{Factor Structure} 

To start, we explore the role of the number and types of factors used in HDF in the predictive performance of the model. Recall that HDF allows for three types of factors: global factors, category-specific factors, and individual-specific factors. To determine a good configuration of the number of each of these factor types, we used predictive performance. Specifically, we estimate the model multiple times, varying the numbers and types of factors, and assess performance by computing the average accuracy, precision, and recall of choice prediction in the holdout year (quarters 45-48). Exhaustively testing every possible combination of numbers of each latent factor type is computationally expensive. Thus, we proceed sequentially: we start by determining the optimal number of \textit{global} factors, as these are the most parsimonious of the three types. We then see if adding additional category and individual factors on top of the optimal number of global factors adds predictive power.

The results are displayed in \Cref{tab:combined_factors_performance}. In the top half of the table, we show the results of changing the number of global factors. We see that the number of latent factors does not dramatically affect holdout performance, though raising the number of factors beyond four leads to (slight) overfitting. Thus, we use four global factors. Next, we consider adding additional individual and category-specific factors on top of those four global factors.\footnote{The sequential selection of global factors and then individual/category factors potentially biases us towards finding global factors. We view this as desirable: global factors are simpler, and thus, by Occam's razor, explaining variability through them is preferable.} The results are displayed in the lower half of \Cref{tab:combined_factors_performance}. Again, we find the addition of these factors offers essentially no out-of-sample gains: accuracy and recall tend to decrease with the addition of any additional factors, while precision modestly increases with the addition of up to two category factors, before dropping. 
 
\begin{table}[ht]
  \centering
  \begin{tabular}{cccccc}
  \toprule
  Global & Category & Indiv. & \multicolumn{3}{c}{Performance Metrics} \\
  Factors & Factors  & Factors  & Accuracy & Precision & Recall \\
  \midrule
  2 & 0 & 0 & 0.6006 & 0.5892 & 0.5208 \\
  \rowcolor{gray!6}3 & 0 & 0 & 0.6034 & \textbf{0.5911} & 0.5260 \\
  4 & 0 & 0 & \textbf{0.6085} & 0.5879 & \textbf{0.5381} \\
  \rowcolor{gray!6}5 & 0 & 0 & 0.6002 & 0.5793 & 0.5302 \\
  7 & 0 & 0 & 0.5987 & 0.5901 & 0.5158 \\
  \midrule
  \rowcolor{gray!6}4 & 0 & 0 & \textbf{0.6086} & 0.5879 & \textbf{0.5382} \\
  4 & 1 & 0 & 0.6003 & 0.5910 & 0.5235 \\
  \rowcolor{gray!6}4 & 2 & 0 & 0.5980 & \textbf{0.6018} & 0.5160 \\
  4 & 3 & 0 & 0.5736 & 0.5621 & 0.5026 \\
  \rowcolor{gray!6}4 & 0 & 1 & 0.6043 & 0.5939 & 0.5218 \\
  4 & 0 & 2 & 0.5996 & 0.5900 & 0.5214 \\
  \rowcolor{gray!6}4 & 0 & 3 & 0.5942 & 0.5779 & 0.5202 \\
  4 & 1 & 1 & 0.5738 & 0.5477 & 0.4981 \\
  \bottomrule
  \end{tabular}
  \caption{Combined table showing the average performance of the HDF model with differing numbers of global, category, and individual-level factors. Bold numbers are the best performing models for each statistic, in each analysis.}
  \label{tab:combined_factors_performance}
\end{table}

That a relatively simple four global factor model achieves the best predictive performance has important substantive implications: it suggests that consumer sensitivity evolution can be explained by global trends. We find no evidence of important category-specific trends or individual-specific trends, at least in terms of forecasting future choices. While individuals' weights on these global factors do vary, isolating the global trends is enough to explain future purchasing. This is important not only because it dramatically simplifies our model of sensitivity evolution, but because it suggests that retailers should focus their marketing research efforts on characterizing broad market trends, rather than category- or individual-specific (e.g., segment-based) trends. 

\paragraph{Benchmarks} Next, we compare the predictive performance of HDF using four global factors against two benchmarks: 
\begin{itemize}
  \item \textbf{GPDH:} Specifically, we use the GP-GPDH variant of GPDH from \cite{dew2020modeling}, which is the GPDH specification with a GP mean function. We focus on GP models for simplicity and to provide reasonable comparisons across the models.
  \item \textbf{Correlated Mixed Logit (CMF):} This benchmark is the mixed logit model with a full covariance matrix across sensitivity parameters, equivalent to the HDF model with all factor weights set to zero. 
\end{itemize}
We focus on these two benchmarks, as they each capture one of the two features of HDF: GPDH captures dynamic heterogeneity, but not correlations across parameters; CMF captures correlations, but not dynamics. For this comparison, we consider not just average fit statistics, but also category-by-category results. We present the full set of statistics in \Cref{table:performance_metrics_by_category}.

\begin{table}[hp]
  \centering
  \begin{tabular}{llccc}
  \hline
  Category & Metric & CML & GPDH & HDF (4) \\
  \hline 
  Overall & Accuracy & 0.5781 & 0.5966 & \textbf{0.6086} \\
          & Precision & 0.5704 & 0.5851 & \textbf{0.5879} \\
          & Recall & 0.4809 & 0.5091 & \textbf{0.5382} \\
  \hline
  \rowcolor{gray!6} Coffee & Accuracy  & 0.6171 & 0.655 & \textbf{0.6618} \\
  \rowcolor{gray!6}  & Precision & 0.6225 & \textbf{0.6559} & 0.6544 \\
  \rowcolor{gray!6}  & Recall    & 0.5686 & 0.6141 & \textbf{0.6285} \\
        Paper Towels & Accuracy  & 0.6787 & 0.7132 & \textbf{0.7168} \\
                     & Precision & 0.6014 & 0.6484 & \textbf{0.652} \\
                     & Recall    & 0.6116 & \textbf{0.6359} & 0.6314 \\
  \rowcolor{gray!6} Soda   & Accuracy  & 0.5643 & 0.5879 & \textbf{0.5942} \\
  \rowcolor{gray!6}  & Precision & 0.5357 & \textbf{0.5507} & 0.5448 \\
  \rowcolor{gray!6}  & Recall    & 0.4271 & 0.4532 & \textbf{0.4769} \\
       Toilet Paper  & Accuracy  & 0.671  & 0.7092 & \textbf{0.7217} \\
                     & Precision & 0.6385 & 0.6816 & \textbf{0.6901} \\
                     & Recall    & 0.6363 & 0.6594 & \textbf{0.6749} \\
  \rowcolor{gray!6} Cereal & Accuracy  & 0.4876 & 0.4898 & \textbf{0.5107} \\
  \rowcolor{gray!6}      & Precision & 0.4369 & 0.4414 & \textbf{0.4589} \\
  \rowcolor{gray!6}      & Recall    & 0.3294 & 0.3564 & \textbf{0.3966} \\
             Mustard & Accuracy  & 0.6350  & 0.6242 & \textbf{0.6444} \\
                     & Precision & 0.6043 & \textbf{0.6098} & 0.6069 \\
                     & Recall    & 0.5750  & \textbf{0.5759} & 0.5942 \\
  \rowcolor{gray!6} Ketchup & Accuracy  & 0.7970  & 0.7892 & \textbf{0.8021} \\
  \rowcolor{gray!6}   & Precision & 0.7524 & 0.741  & \textbf{0.7584} \\
  \rowcolor{gray!6}   & Recall    & 0.7354 & 0.7342 & \textbf{0.7465} \\
  \hline
  \end{tabular}
  \caption{Comparing the HDF, GPDH, and CML models, category-by-category.}
  \label{table:performance_metrics_by_category}
\end{table}

Overall, we see modest gains of HDF over GPDH, and both HDF and GPDH over CML. That both dynamic models outperform the static CML suggests that, across categories, dynamics are important. However, the degree to which dynamics matter, and to which HDF outperforms GPDH at capturing those dynamics, varies by category. The most dramatic gains in terms of HDF over the static CML were in toilet paper (gain of 5\% accuracy), coffee (4.5\%), and paper towels (4\%). This suggests that individuals' marketing sensitivities evolved most dramatically in these three categories. On the other hand, the most dramatic gains over GPDH were in cereal and mustard (gains of 2\% accuracy), followed by ketchup and toilet paper (1.3\%). 

That the HDF with just four factors consistently outperforms GPDH is, in itself, surprising and interesting: GPDH allows for flexible, individual-specific trajectories, whereas HDF assumes that all dynamic heterogeneity is a result of a small number of global trends. Interpreted substantively, these results suggest that much of the dynamics in individuals' preferences can be explained a small number of latent trends, and that forecasting these trends is essential for forecasting changes in preferences. 

\begin{figure}
    \centering
    \includegraphics[width=\textwidth]{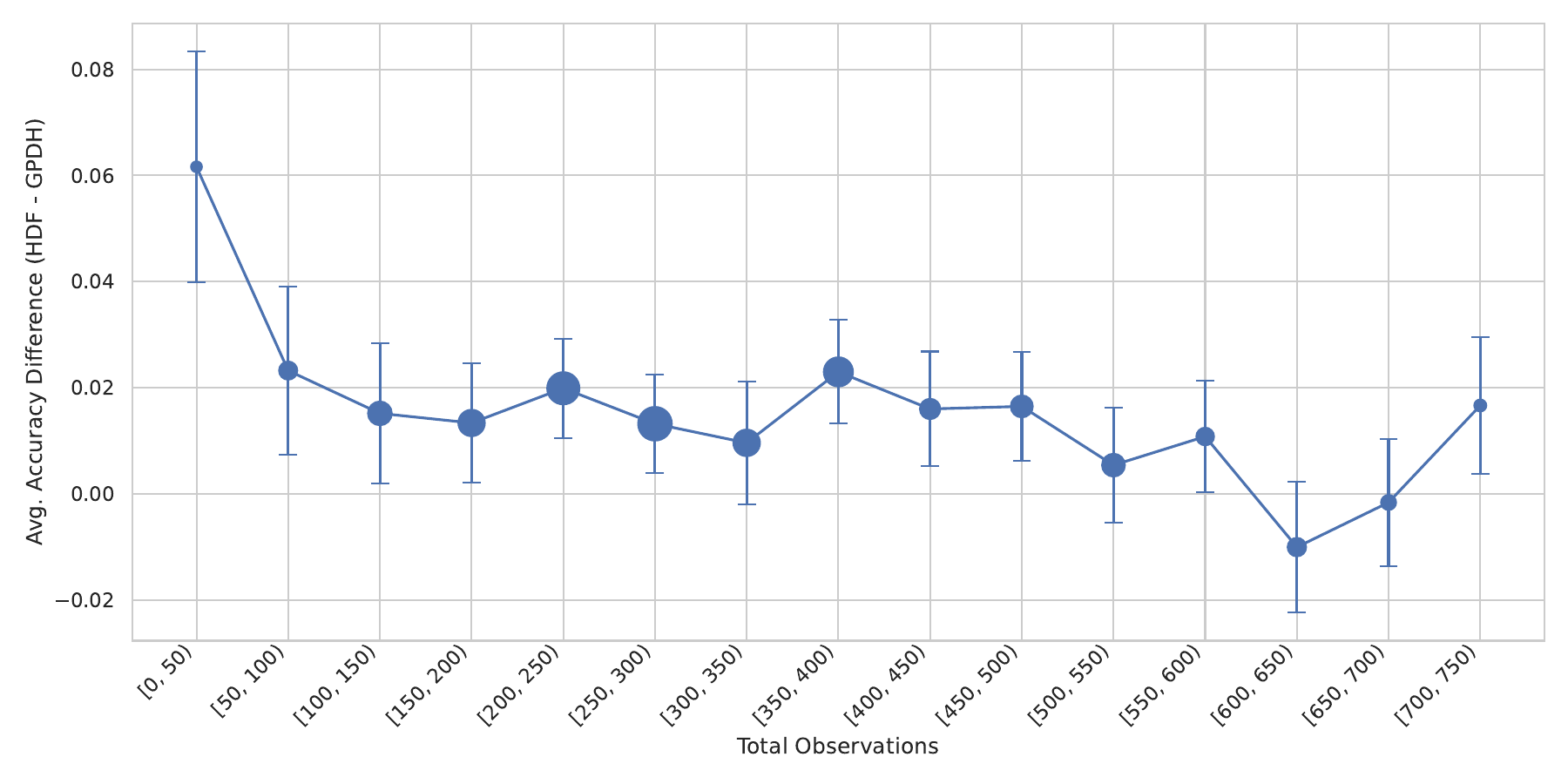}
    \caption{The average difference in test accuracies between HDF and GPDH, for differing numbers of training observations, binned in intervals of 50. The observations being averaged are category-individual pairs (i.e., person $i$ purchasing in category $c$). The size of the point represents the number of observations in each bin. Error bars are standard errors.}
    \label{fig:accuracy-diff-by-obs}
\end{figure}

We can also examine predictive performance at the individual level. Beyond imposing a factor structure on the dynamics, HDF also allows for sharing of information across categories. This sharing means that a paucity of data in one category can be supplemented by a richness in another when making predictions. For this reason, we should see the strongest performance gains for individuals who have relatively little data. Indeed, we do: in \Cref{fig:accuracy-diff-by-obs}, we show how the average difference in test accuracies between HDF and GPDH varies as a function of the number of total training observations for the individual. Notably, there is a large spike at the beginning, for individuals with fewer than 50 training observations. Then, the accuracy difference declines, hovering between 0-2\% for individuals with more than 100 observations. 

\subsection{Factors and Correlations} \label{sec:param-estimates}

Now, we turn our attention to the parameter estimates from the HDF model with four global factors. Specifically, we explore a series of related questions: what are the four factors? Can they be meaningfully interpreted? To what extent do parameters exhibit correlated dynamics? And what do those correlations tell us about evolution of marketing mix sensitivities during the period examined?\footnote{The substantive insights in this section are subject to the caveat about endogeneity mentioned previously. For instance, it is possible that one common factor may reflect endogeneity bias, or more generally that the factor structure may change after accounting for it.}

\begin{figure}
  \centering
  \includegraphics*[width=0.7\textwidth]{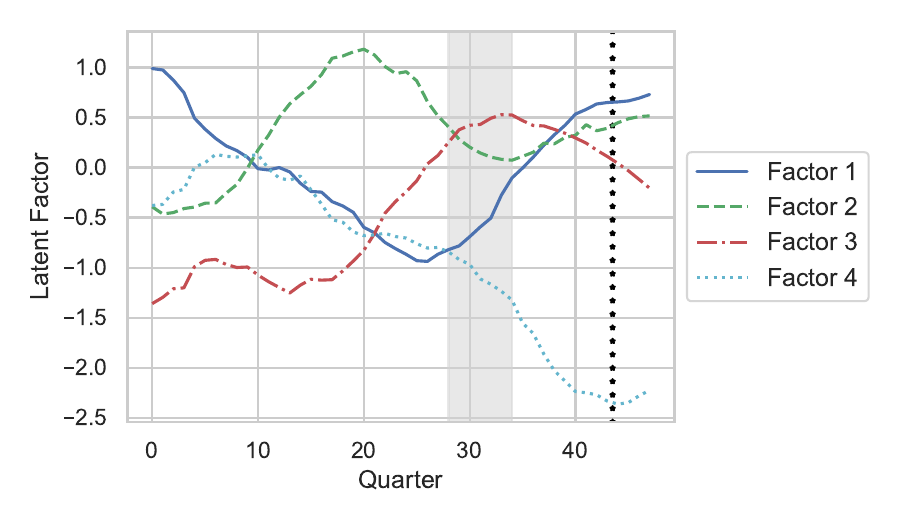}
  \caption{Posterior means of the four latent factors. The grey shaded region indicates the period of the 2008 Great Recession. The black starred line indicates the end of the calibration period.}
  \label{fig:four_factors}
\end{figure}

\paragraph{Latent Factors} In \Cref{fig:four_factors}, we show the posterior mean estimates of the four latent factors. We see that, first, all appear quite smooth. This smoothness suggests that the implicit regularization through the prior on $\rho$ is working. In terms of the actual dynamics captured by these factors, we see many interesting effects around the time of the Great Recession (the shaded rectangle). For instance, Factor 1, which had been decreasing, suddenly starts increasing in that window. Factor 4, on the other hand, which had been steadily decreasing, accelerated its decrease after the recession. Factors 2 and 3 are a bit more ambiguous, but both seemed to level off around the time of the recession. That being said, directly interpreting the factors is difficult: there is significant heterogeneity in weights on each factor, even for a single parameter in a single category. Moreover, the weights are not sparse, meaning individual customers often have non-negligible weight on \textit{all} four factors. This makes it difficult to cleanly interpret the factors separately from one another.

\paragraph{Correlated Dynamics} Although the factors themselves are difficult to interpret, the weight correlation matrix $\Lambda_\omega$ is directly interpretable. Recall that this matrix captures correlations of individual-level factor weights across different parameters (through Equations \ref{eq:weight_vector_prior} and \ref{eq:corr_mat_weights}). For example, if element $(s,s')$ of $\Lambda_\omega$ is positive, it means that individuals' weights on the latent factors in determining parameter $s$ are positively correlated with their weights for determining parameter $s'$. The larger the correlation estimate, the more those two preference sensitivities rely on the same latent factors, on average across people. Thus, these correlation estimates capture how similar two preference trajectories tend to be, and we thus refer to them as estimates of correlated dynamics. 

\begin{figure}
  \centering
  \includegraphics[width=0.7\textwidth]{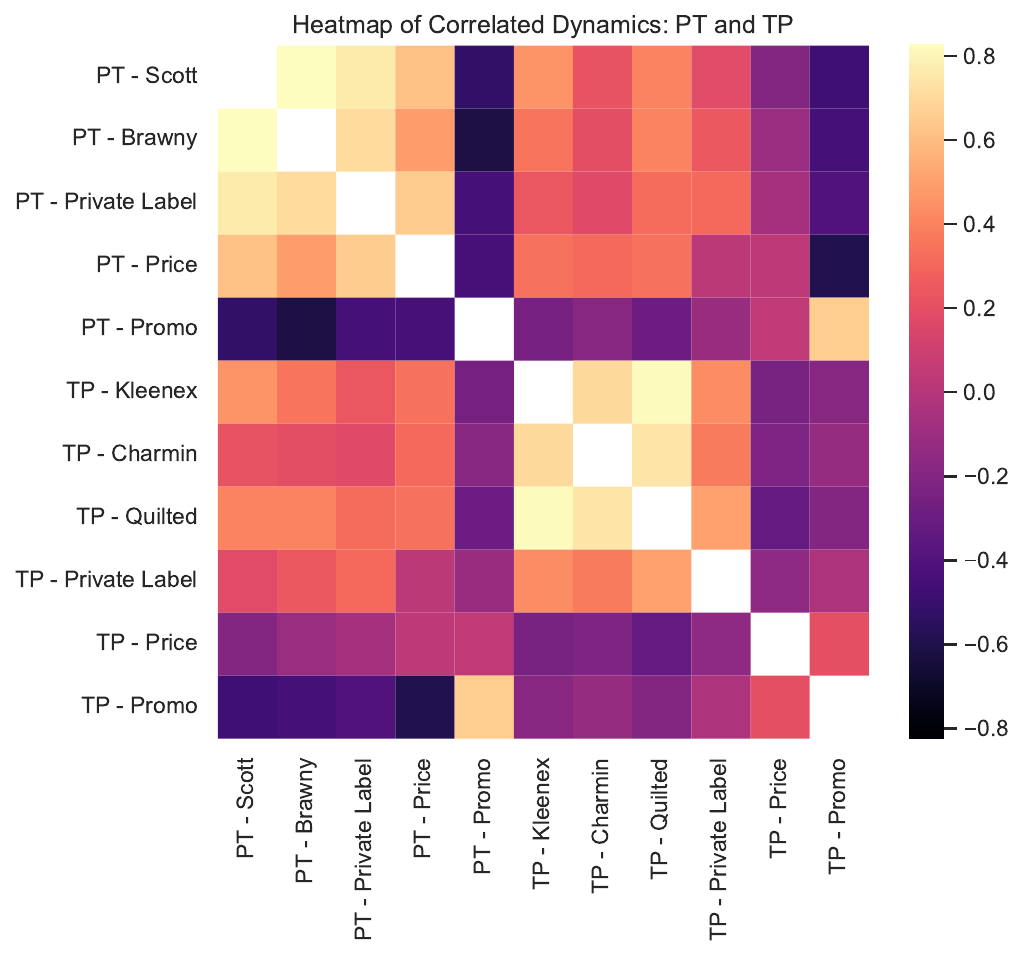}
  \caption{Heatmap of the submatrix of $\Lambda_\omega$ corresponding to parameters in the PT (Paper Towels) and TP (Toilet Paper) categories. Brigher colors indicate more positive correlation, darker colors indicate more negative correlation.}
  \label{fig:pt-tp-heatmap}
\end{figure}

From the posterior mean estimate of $\Lambda_\omega$, we find considerable correlations in preference dynamics, both within and across categories. Examining the full correlation matrix is unwieldy; hence, we focus on several submatrices of interest, and include a visualization of the full correlation matrix in \Cref{ap:full-corr}. As a motivating example, in \Cref{fig:pt-tp-heatmap}, we plot the submatrix of $\Lambda_\omega$ corresponding to parameters in the paper towels (PT) and toilet paper (TP) categories. We see that, within each category, there are blocks of high positive correlation among the brand intercepts, suggesting changes in brand preferences tend to be correlated. These correlations could be driven by changes in preference for the reference brand, or by correlated dynamics in preferences for the non-reference brands. Perhaps more interesting are the sizable positive correlations \textit{across} categories, captured by the off-diagonal blocks, which suggest that brand preference evolution in toilet paper is related to paper towels, and vice versa. We also see a strong positive correlation between the two categories' promotion sensitivities, which are negatively correlated with the other parameters. This negative correlation suggests that the dynamic factors determining these two promotion sensitivities are different from those determining the other parameters.

More broadly, there are positively correlated dynamics between the brand intercepts of many categories. In \Cref{tab:avg-brand-cors}, we show the average values of $\Lambda_\omega$ between each category's brand intercepts, capturing how correlated the dynamics are between these groups of parameters, on average. First, on the diagonal, we see that all categories exhibit high within-category correlations. Across categories, there is more variation: paper towels, toilet paper, mustard, and ketchup all exhibit high cross-category correlations, while coffee, soda, and cereal exhibit relatively low correlations. Coffee, in particular, exhibits almost no cross-category correlation. This pattern suggests that understanding how brand preferences are evolving in paper towels, toilet paper, mustard, or ketchup is informative about how brand preferences are evolving in the others, whereas coffee's brand preference dynamics operate more independently. One possible explanation is the level of differentiation within these product categories: paper towels, toilet paper, mustard, and ketchup are all relatively homogeneous categories, where the store brand may play an outsized role. On the other hand, coffee, soda, and cereal are more differentiated. In more differentiated categories, consumers may have more idiosyncratic preferences, leading to less correlated dynamics.

\begin{table}
  \centering
  \begin{tabular}{lccccccc}
  \toprule
   & Coffee & PT & Soda & TP & Cereal & Must. & Ket. \\
  \midrule
  Coffee & 0.201 & \textcolor{gray!50}{-0.013} & \textcolor{gray!50}{0.009} & -0.043 & 0.016 & \textcolor{gray!50}{0.002} & 0.041 \\
  PT & \textcolor{gray!50}{-0.013} & 0.574 & 0.087 & 0.216 & 0.057 & 0.236 & 0.112 \\
  Soda & \textcolor{gray!50}{0.010} & 0.097 & 0.300 & 0.083 & 0.035 & 0.040 & 0.093 \\
  TP & -0.046 & 0.230 & 0.080 & 0.475 & 0.045 & 0.172 & 0.070 \\
  Cereal & 0.018 & 0.063 & 0.035 & 0.046 & 0.229 & 0.036 & 0.038 \\
  Mustard & \textcolor{gray!50}{0.002} & 0.209 & 0.032 & 0.143 & 0.029 & 0.251 & 0.138 \\
  Ketchup & 0.036 & 0.100 & 0.075 & 0.058 & 0.030 & 0.138 & 0.320 \\
  \bottomrule
  \end{tabular}
  \caption{Average correlations between each category's brand intercept parameters. Grey values indicate average correlations whose 95\% credible intervals include zero. ``PT'' stands for Paper Towels; ``TP'' stands for Toilet Paper.}
  \label{tab:avg-brand-cors}
\end{table}

In the paper towels and toilet paper example, we noted the high cross-category correlation of promotion sensitivity dynamics. This raises the question: to what extent are dynamics in the same marketing mix sensitivities correlated across categories more generally? Recall that the literature suggests that the macroeconomic environment (and especially a recession) tends to broadly affect consumer-level quantities like price elasticity and demand for private label goods. Through HDF, we can examine those quantities from a slightly different perspective: how related were the dynamics in these types of individual-level sensitivity parameters across categories? 

In our model, there are three features shared across all categories: price, promotion, and private label brand. We find that there are correlated dynamics between the vast majority of these parameters: for private label intercepts and promotion sensitivities, all but one entry of the corresponding submatrix of $\Lambda_\omega$ are positive, with credible intervals all excluding zero. For price sensitivities, of the 21 correlation parameters, three are significant and negative, and three are not significantly different from zero, while the rest are positive and significant. In terms of magnitude, the highest correlated dynamics are between promotion sensitivites (mean 0.289), followed by private label intercepts (mean 0.181), and finally price sensitivities (mean 0.124).\footnote{We include the full correlation matrices for each of these parameters in \Cref{ap:price-promo-corr-submats}.} Within each parameter type, there are also noteworthy findings. For instance, we see very correlated dynamics between the promotion sensitivities of paper towels, coffee, toilet paper, and ketchup (all correlated above 0.5), more so than any other features in any other categories. The ability of HDF to capture these kinds of a priori unexpected connections between categories highlights the flexibility and utility of the model. 

It is worth noting that, while these patterns echo some of the findings in the literature, they are fundamentally different: here, we are looking at \textit{changes} in preference sensitivities, not just levels. To make the contrast clear, consider our contribution relative to \cite{ainslie1998similarities}. They find that sensitivities to price and feature/promotion are correlated across categories within households, and that the correlation was larger for promotion. At face value, this finding is similar to ours. However, their finding is specific to one point in time. Our findings in this section show that, when looked at over long time horizons, the way these marketing sensitivities \textit{change} is also correlated across parameters and categories, sometimes in surprising ways. Moreover, that evolution can be explained using a relatively small number of global dynamic factors.

\subsection{Implications: Dynamic Heterogeneity and Elasticities} \label{sec:capturing-dyn-het}

Until now, we have focused exclusively on population-level quantities. In this section, we consider the individual-level insights enabled by the HDF model. We start by examining what patterns of dynamic heterogeneity are captured by the HDF specification. We then show how patterns of dynamic heterogeneity translate into important managerial quantities like price elasticity, and highlight the more precise understanding of preference evolution enabled by HDF.

\begin{figure}
  \centering
  \includegraphics[width=\textwidth]{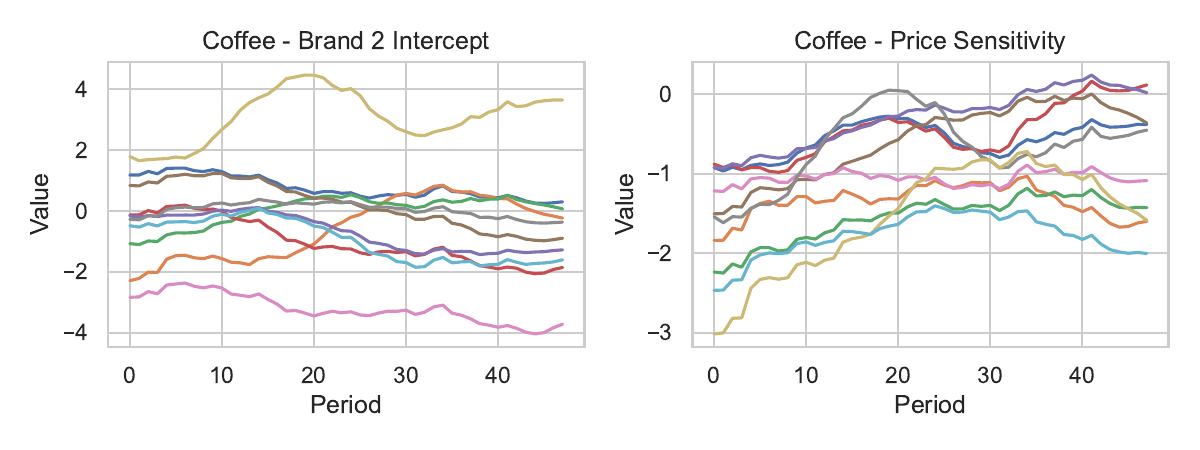}
  \caption{A sample of 10 random individual-level parameter trajectories from the coffee category. At left are trajectories for brand 2 intercepts; at right are price sensitivity trajectories.}
  \label{fig:ind_price_trajectories}
\end{figure}

First, to illustrate the range of individual-level preference trajectories captured by HDF, in \Cref{fig:ind_price_trajectories}, we plot brand intercept and price sensitivity trajectories from the coffee category, for 10 randomly selected customers. We see there is substantial dynamic heterogeneity: consistent with the simulations discussed previously, despite the fact that the trajectories are generated solely through heterogeneous intercepts and weights on the same four latent factors, we see heterogeneous trajectories in both cases, especially for the brand 2 intercept. We also see that, for price sensitivity, there is a common average trend upwards. While there are still individual variations around that trend, it seems that coffee customers' price sensitivity generally decreased during the observation window. Such a population trend is not explicitly part of the  HDF specification, as it is in GPDH, but common trajectories can still emerge. 

\paragraph{Elasticities} While sensitivities alone can be difficult to interpret, they are an essential ingredient in computing elasticities, the bedrock of most marketing mix optimization. As the HDF likelihood is just a multinomial logit, computing elasticities is straighforward:
\begin{equation}
  E_{ijt}^\mathrm{x} = \beta_{i p_x}(t) \times x_{ijt} \times [1 - \pi_{ijt}],
\end{equation}
where $x$ is a specific feature (e.g., price), $p_x$ denotes the parameter index corresponding to that feature's sensitivity, and $\pi_{ijt}$ is the probability that person $i$ purchases brand $j$ at time $t$.\footnote{Throughout this section, we will maintain the negative sign on the price elasticity. Thus, an ``increase'' in price elasticity means individuals are becoming more price inelastic. Likewise, a ``decrease'' in price elasticity means individuals are becoming more price elastic.} 

\begin{figure}
  \begin{adjustbox}{center}
  \includegraphics[width=1.1\linewidth]{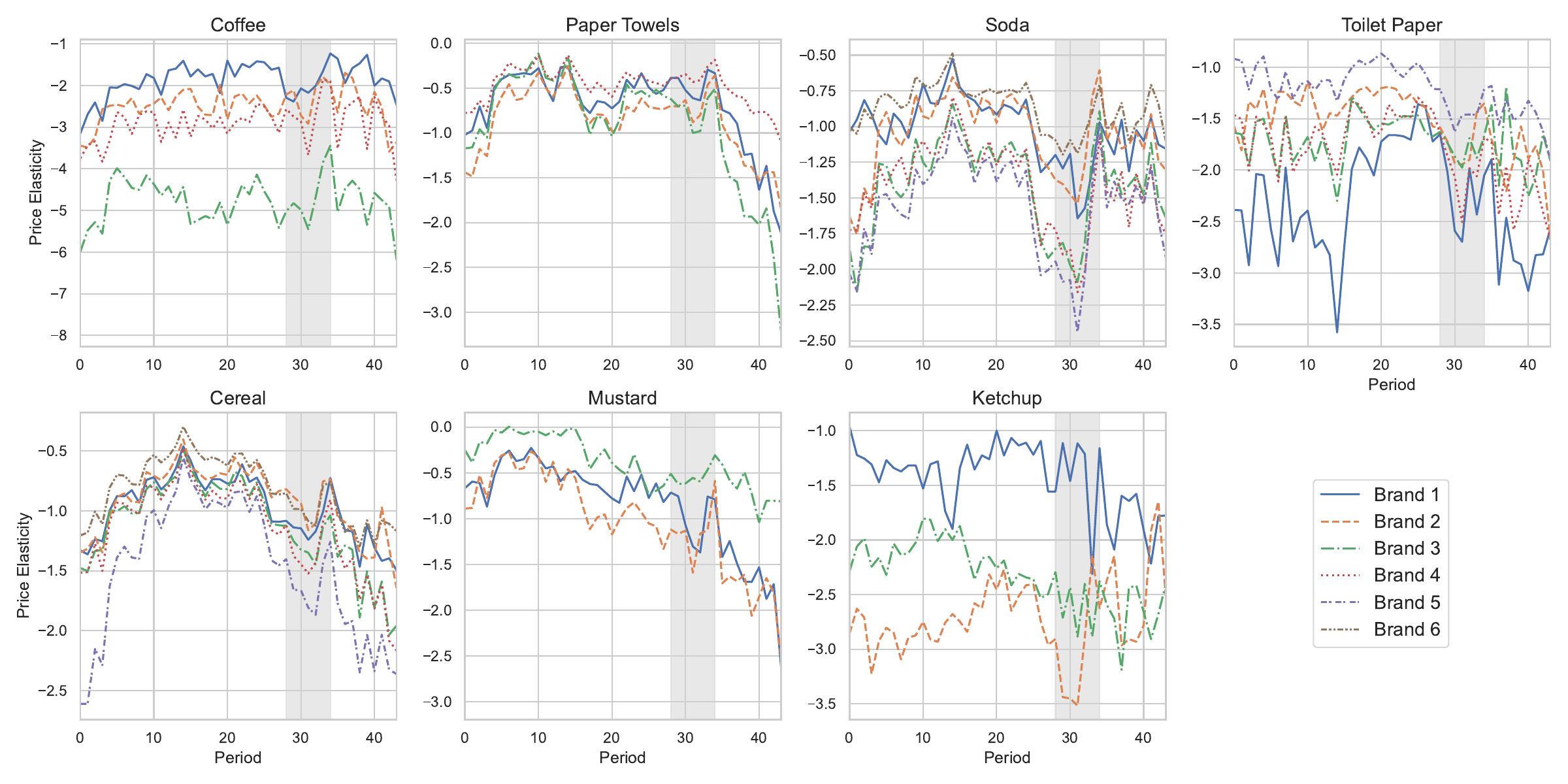}
  \end{adjustbox}
  \caption{Category-level own-price elasticity dynamics, averaged over individuals. Each panel represents a category, each line a brand within the category. The grey vertical area marks the start and end of the Great Recession.}
  \label{fig:cat-price-elas}
\end{figure}

In Figure \ref{fig:cat-price-elas}, we plot the average price elasticity across individuals, broken out by categories. We see there are obvious dynamics, just as there were in the parameter estimates. Moreover, many of these dynamics are centered around the recession, demarcated on the plot by the grey box. For instance, in paper towels, there is a stark decreasing trend coinciding with the end of the recession, suggesting consumers became much more price elastic in the post-recession period. In soda, we see a very different pattern, with a striking decrease in the few quarters before the official recession dates, and then a bounce back at the end, suggesting that changes to price elasticity during the downturn were relatively short-lived. A similar pattern can be observed in other categories, like toilet paper and ketchup (brand 2). Cereal also exhibits a decrease just before the official recession, though it continues modestly decreasing in the post-recession period. The fact that many of these elasticity curves begin decreasing \textit{before} the official recession suggests that consumers may have experienced changing economic circumstances, or more generally the effects of the macroeconomic environment sooner than the official economic contraction. The findings are also reminiscent of the findings of \cite{dube2018income}, who noted that some changes in consumer preferences --- like demand for private label products --- began before the recession.

\begin{figure}
  \centering
  \includegraphics[width=0.6\textwidth]{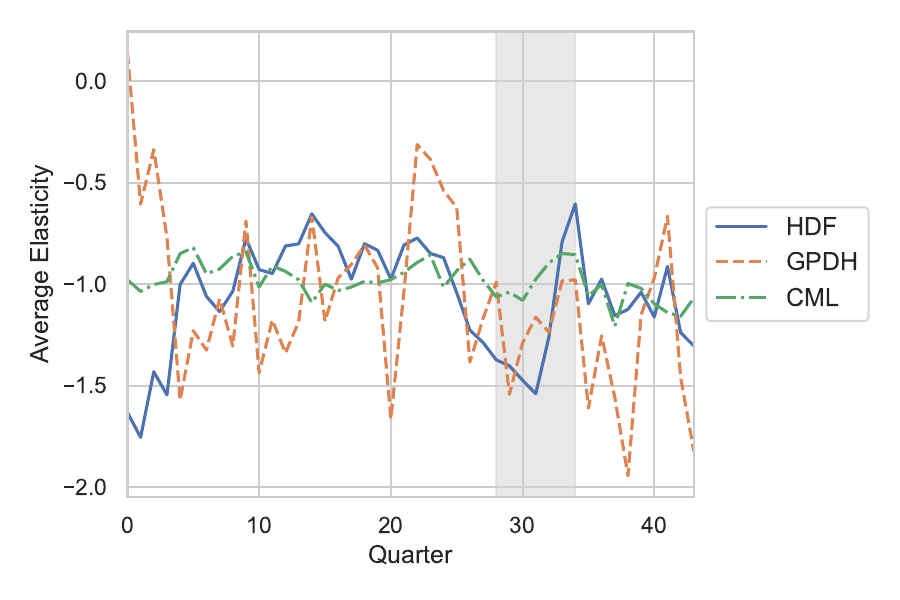}
  \caption{Average own-price elasticity over time for Pepsi, estimated by HDF, GPDH, and CMF, }
  \label{fig:avg-elas-across-models}
\end{figure}

To understand the value of HDF in understanding these dynamics, we now compare the average elasticities learned from it versus GPDH and the static CMF. To do so, we focus on an especially interesting product category, soda, which, as we saw in \Cref{fig:cat-price-elas}, exhibited strong dynamics during the recession. To compare these elasticities to GPDH and CMF, we pick a single brand -- Brand 2, Pepsi -- which exemplified those dynamics. In \Cref{fig:avg-elas-across-models}, we show the Pepsi elasticity estimates from HDF, GPDH, and CMF. There are several noteworthy findings: first, as described in \cite{dew2020modeling}, the static model exhibits a mild attenuation bias, meaning it is biased toward zero. Second, the variance in the GPDH estimate is quite high, which masks the interesting dynamics noted in the HDF results. This pattern is not just true for this brand: across brands and parameters, we find that the HDF elasticity estimates exhibit lower variance, and more clearly interpretable trends, relative to GPDH. In turn, these more stable trends enable researchers to better understand how preferences have changed over time, as in the soda example. 

\begin{figure}
  \centering
  \begin{adjustbox}{center}
  \includegraphics[width=1.1\textwidth]{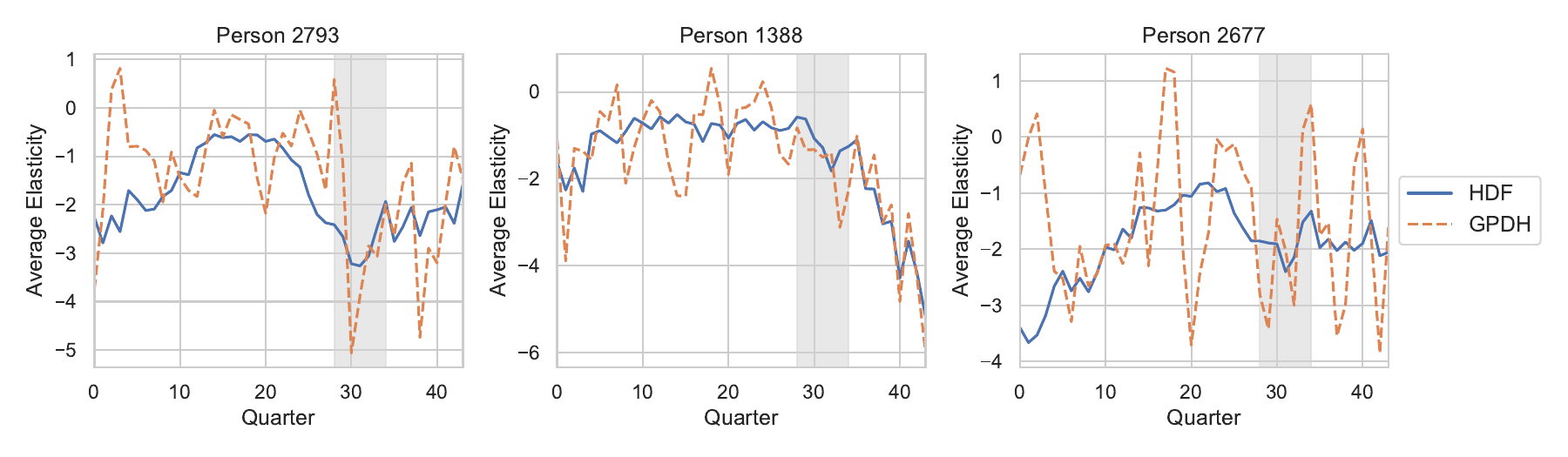}
  \end{adjustbox}
  \caption{Three selected individuals and their Pepsi own-price elasticity curves, estimated by HDF and GPDH.}
  \label{fig:ind-elas-across-models}
\end{figure}

Just as there are obvious dynamics on average, there are also obvious dynamics at the individual-level.  In \Cref{fig:ind-elas-across-models}, we plot the same brand's own-price elasticity curves for three individuals. These individuals purchased in the category in every period, allowing us to compute the elasticity over time. We can see the similar differences between the three models: the GPDH estimates exhibit high variance, whereas the HDF estimates are much more smooth over time. Crucially, this difference has direct implications for interpretation, with trajectories emerging much more prominently under HDF. HDF is also much less prone to yielding nonsensical positive elasticities. Taken together, these examples highlight the potential of HDF to aid in the analysis of dynamic effects. 

\begin{figure}
    \begin{adjustbox}{center}
        \includegraphics[width=1.1\textwidth]{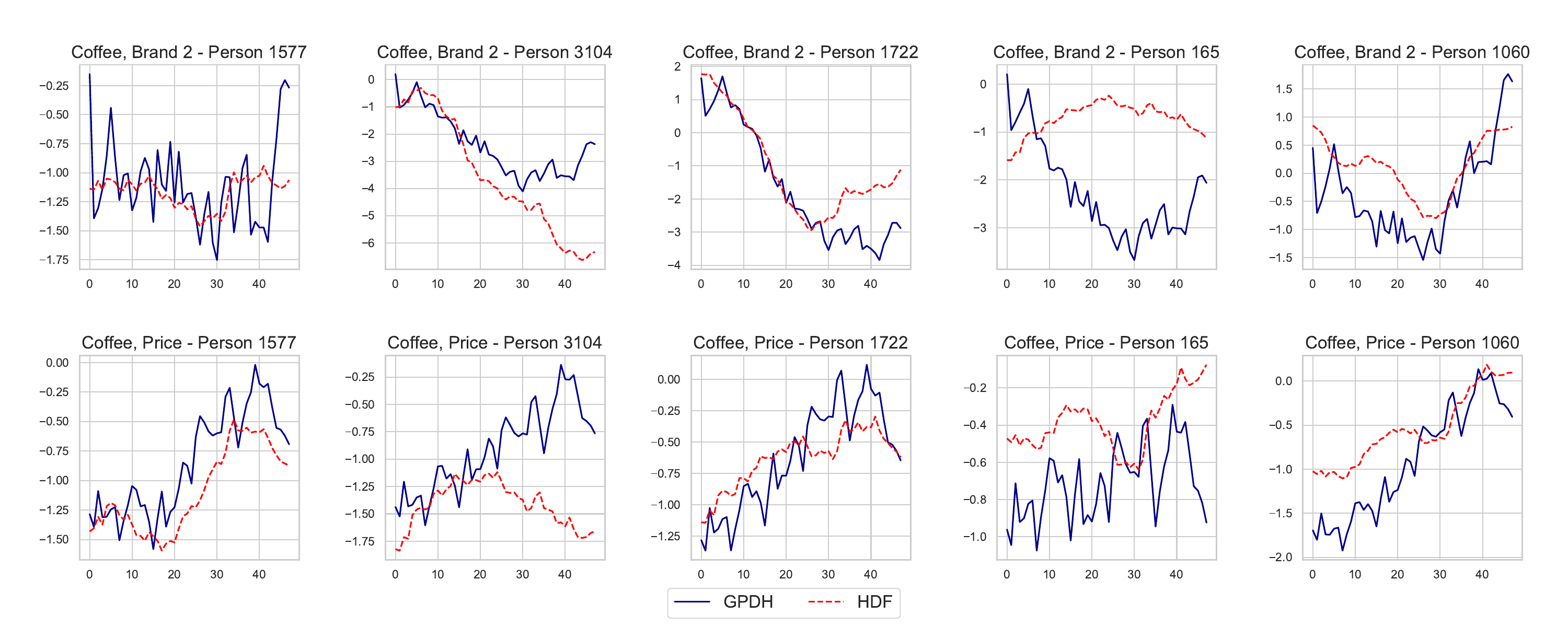}
    \end{adjustbox}
    \caption{Comparison of parameter trajectories for Coffee's brand 2 intercept and price sensitivity for five randomly selected individuals. The solid line is GPDH; the dashed line is HDF.}
    \label{fig:coffee-gpdh-vs-hdf}
\end{figure}

\paragraph{Why does HDF yield more stable estimates?} That HDF yields more stable, efficient estimates than GPDH is no accident. Rather, this lower variance stems from the same mechanisms that enable its strong predictive performance: high regularization through the latent factors, and the partial pooling of information enabled by the correlation structure across parameters. These two forces lead to smoother, more reliable individual-level curves, and in turn, to more reliable aggregate insights. To see these effects more clearly, we revisit the analysis from \Cref{fig:coffee-gpdh-vs-hdf}. In \Cref{fig:coffee-gpdh-vs-hdf}, we take those same two parameters --- coffee's brand 2 intercept, and its price sensitivity --- and compare the estimated sensitivity trajectories for five randomly selected customers. In many cases, HDF captures the general trajectory of GPDH, but not always. In all cases, however, we see that the GPDH estimates are much higher variance than the HDF estimates. 

There are two potential explanations for the cases where HDF deviates strongly from GPDH. First, GPDH may be overfitting those trajectories; or second, the 4-factor solution of HDF may not be sufficiently flexible to capture that particular pattern of variation. While our predictive validation suggests the former, we can also consider the latter, which raises the question: how restrictive really is HDF? To answer that question more broadly, we ran several further analyses. In each of these analyses, we assumed the GPDH is the ``ground truth'' dynamic heterogeneity, and then examined to what extent HDF was able to capture that ground truth. Specifically, we computed two statistics: the correlations between the two models' parameter trajectories,\footnote{i.e., $\mathrm{cor}(\beta_{is}^\mathrm{GPDH}(1, \ldots, T), \beta_{is}^\mathrm{HDF}(1,\ldots,T))$)} and the McFadden pseudo-$R^2$ of HDF relative to GPDH, taking the CML model as the null model.\footnote{In this case, McFadden's pseudo-$R^2$ is defined as $R^2 = \frac{1 - L_\mathrm{HDF}/L_\mathrm{CML}}{1 - L_\mathrm{GPDH}/L_\mathrm{CML}}$,
where $L_m$ is the log-likelihood of model $m$. We use the CML as the null model, as it is nested in both approaches, by disabling the dynamics.}
Across the two statistics, we find the same result: the four factor HDF model can explain a substantial portion of the variance captured by GPDH. The average correlation between the two models' sensitivities is 0.29, with 72\% of trajectories having a positive correlation, and 39\% having a correlation above 0.5. The pseudo-$R^2$, which more directly captures the notion of variance in GPDH explained by HDF, is 0.66. This finding echoes many of our earlier results: dynamics in customer-level marketing sensitivities  can be explained by a few global factors. When combined with the predictive results, and the increased efficiency documented above, these results suggest that the simpler and more efficient model structure of HDF does not come at a substantial cost, in terms of flexibility, but does offer substantial benefits.

\section{Conclusion} \label{sec:conclusion}

In this work, we develop a flexible framework --- the hierarchical dynamic factor (HDF) model --- to capture cross-category, dynamic, individual-level marketing sensitivities. HDF leverages latent dynamic factors, modeled through Gaussian processes, and a hierarchical correlation structure across customers to accurately infer dynamics in customers' sensitivities to marketing variables. These latent factors parsimoniously capture common trends in markets, and inferring them in a fully Bayesian fashion allows for the sharing of information across product categories. Compared to extant dynamic heterogeneity specifications, HDF provides more precise estimates that can be used for the optimization of the marketing mix. 

Our work makes both methodological and substantive contributions. Methodologically, our HDF specification is new, and presents a simpler way of capturing dynamic heterogeneity, while accounting for correlated dynamics across parameters. To our knowledge, it is the first adaptation of multi-output Gaussian processes to marketing problems. This specification also outperforms previously developed brand choice models in various forecasting metrics, and achieves more reliable price elasticity estimates. Substantively, we show that a small number of latent factors can accurately capture preference dynamics, and highlight the surprising patterns of correlations that exist in individual-level dynamics across product categories. These findings suggest there is value in pooling information across categories in terms of predicting the dynamics in markets, and suggest that common trends govern purchasing dynamics across many disparate product categories.

There are several potentially interesting avenues for future research. First, while interesting, our data are also limited: we focus on a subset of the IRI data, where customers were active over our sample period and made purchases at the start and the end of our sample period. Doing so allows us to rule out that the observed dynamic heterogeneity was driven by customer attrition. However, this precludes us from studying more sparse settings where the information sharing of the HDF may even further improve insights and performance. In general, we see considerable room for application of HDF even beyond grocery purchasing data, including to dynamic and sparse settings like e-commerce purchasing. Even within grocery purchases, further applications of HDF to events like the COVID-19 pandemic, or to categories experiencing a disruption, like the introduction of Greek yogurt in the yogurt category, may yield even more interesting dynamic results. 

Second, the HDF specification we proposed is simple, and as we have documented, works well empirically. However, the literature on multi-output GP models is vast, including a wealth of specifications from geostatistical and machine learning applications. Our specification was inspired by perhaps the simplest of the MOGP models: the linear model of coregionalization. Future work may examine more complex specifications for the cross-category correlations, or even allow the cross-category correlation matrix to itself evolve \citep[e.g.,][]{liu2018remarks}. Such generalizations of our framework may further improve our ability to understand correlated, dynamic heterogeneity.

\newpage
\singlespacing
\bibliographystyle{jmr}
\bibliography{mcdh}

\clearpage
\section{Appendix A: Visualization of Full Correlation Matrix} \label{ap:full-corr}

\begin{figure}[h]
  \includegraphics[width=\textwidth]{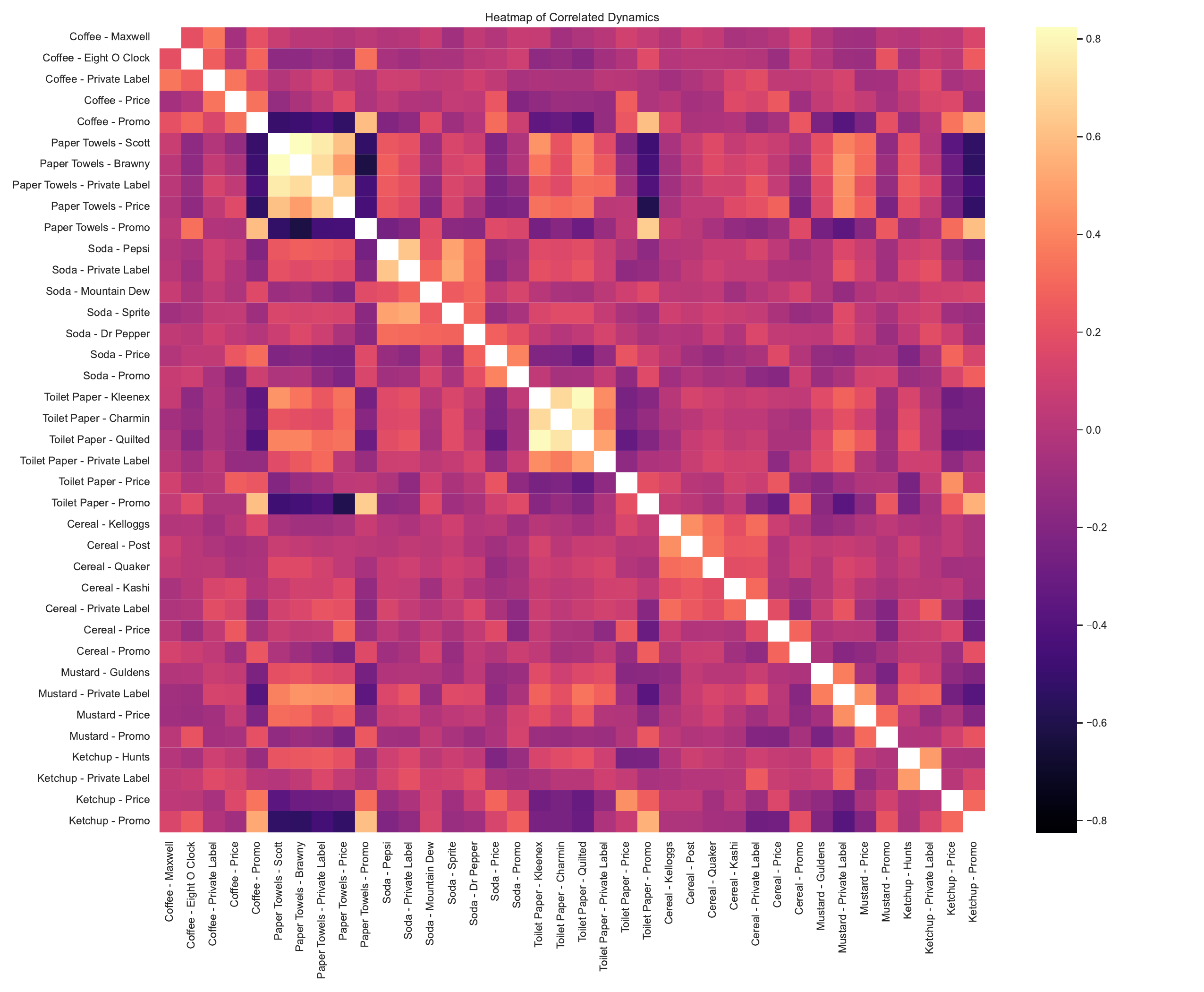}
  \caption{Heatmap of the posterior mean of $\Lambda_\omega$, showing the correlations in the dynamics between all sensitivities in the model.}
\end{figure}

\clearpage
\section{Appendix B: Price, Promotion, and Private Label Correlations} \label{ap:price-promo-corr-submats}

\ifjmr
\renewcommand{\arraystretch}{1}
\fi
\begin{table}[!h]
  \centering
  \resizebox{0.65\textwidth}{!}{
  \begin{tabular}{lccccccc}
  \toprule
   & Coffee & PT & Soda & TP & Cereal & Must. & Ket. \\
  \midrule
  Coffee & 1 & 0.174 & 0.226 & 0.260 & 0.241 & \textcolor{gray!50}{0.055} & 0.156 \\
  PT & 0.174 & 1 & -0.243 & \textcolor{gray!50}{0.028} & 0.278 & 0.266 & -0.258 \\
  Soda & 0.226 & -0.243 & 1 & 0.219 & 0.166 & -0.044 & 0.295 \\
  TP & 0.260 & \textcolor{gray!50}{0.028} & 0.219 & 1 & 0.235 & -0.023 & 0.443 \\
  Cereal & 0.241 & 0.278 & 0.166 & 0.235 & 1 & \textcolor{gray!50}{0.012} & 0.154 \\
  Mustard & \textcolor{gray!50}{0.055} & 0.266 & -0.044 & -0.023 & \textcolor{gray!50}{0.012} & 1 & -0.041 \\
  Ketchup & 0.156 & -0.258 & 0.295 & 0.443 & 0.154 & -0.041 & 1 \\
  \bottomrule
  \end{tabular}
  }
  \caption{Posterior mean estimate of the submatrix of $\Lambda_\omega$ corresponding to price variables. Grey values indicate correlations whose 95\% credible intervals include zero. ``PT'' stands for Paper Towels; ``TP'' stands for Toilet Paper.}
  \label{tab:price-correlation}
\end{table}
\vspace*{\baselineskip}
\begin{table}[!h]
  \centering
  \resizebox{0.65\textwidth}{!}{
  \begin{tabular}{lccccccc}
  \toprule
    & Coffee & PT & Soda & TP & Cereal & Must. & Ket. \\
  \midrule
  Coffee & 1 & 0.594 & 0.084 & 0.605 & 0.226 & 0.200 & 0.525 \\
  PT & 0.594 & 1 & 0.142 & 0.655 & 0.166 & 0.241 & 0.602 \\
  Soda & 0.084 & 0.142 & 1 & 0.170 & 0.079 & 0.117 & 0.266 \\
  TP & 0.605 & 0.655 & 0.170 & 1 & 0.264 & 0.233 & 0.555 \\
  Cereal & 0.226 & 0.166 & 0.079 & 0.264 & 1 & \textcolor{gray!50}{-0.061} & 0.200 \\
  Mustard & 0.200 & 0.241 & 0.117 & 0.233 & \textcolor{gray!50}{-0.061} & 1 & 0.216 \\
  Ketchup & 0.525 & 0.602 & 0.266 & 0.555 & 0.200 & 0.216 & 1 \\
  \bottomrule
  \end{tabular}
  }
  \caption{Posterior mean estimate of the submatrix of $\Lambda_\omega$ corresponding to promotion variables. Grey values indicate correlations whose 95\% credible intervals include zero. ``PT'' stands for Paper Towels; ``TP'' stands for Toilet Paper.}
\end{table}
\vspace*{\baselineskip}
\begin{table}[!h]
  \centering
  \resizebox{0.65\textwidth}{!}{
  \begin{tabular}{lccccccc}
  \toprule
   & Coffee & PT & Soda & TP & Cereal & Must. & Ket. \\
  \midrule
  Coffee & 1 & 0.123 & 0.101 & \textcolor{gray!50}{0.015} & 0.182 & 0.129 & 0.169 \\
  PT & 0.123 & 1 & 0.084 & 0.308 & 0.224 & 0.443 & 0.152 \\
  Soda & 0.101 & 0.084 & 1 & 0.123 & 0.150 & 0.159 & 0.185 \\
  TP & \textcolor{gray!50}{0.015} & 0.308 & 0.123 & 1 & 0.114 & 0.276 & 0.097 \\
  Cereal & 0.182 & 0.224 & 0.150 & 0.114 & 1 & 0.224 & 0.251 \\
  Mustard & 0.129 & 0.443 & 0.159 & 0.276 & 0.224 & 1 & 0.300 \\
  Ketchup & 0.169 & 0.152 & 0.185 & 0.097 & 0.251 & 0.300 & 1 \\
  \bottomrule
  \end{tabular}
  }
  \caption{Correlated dynamics between private label brand intercepts across categories. Grey values indicate correlations whose 95\% credible intervals include zero. ``PT'' stands for Paper Towels; ``TP'' stands for Toilet Paper.}
\end{table}

\end{document}